\documentclass[aps,prl,groupedaddress,twocolumn,notitlepage,superscriptaddress,10pt]{revtex4-2}

\usepackage{pdfpages} % for adding .pdf appendix
\makeatletter
\AtBeginDocument{\let\LS@rot\@undefined}
\makeatother
\usepackage{float}
\usepackage{graphicx}  % needed for figures
\usepackage{xcolor}
\usepackage{bm}        % for math
\usepackage{braket}
\usepackage{amssymb}   % for math
\usepackage{epstopdf}
\usepackage{xfrac}
\usepackage{cancel}
\usepackage{soul}
\usepackage{amsmath}
\usepackage{amsthm}
\usepackage{amsfonts}
\usepackage{bbm}
\definecolor{darkblue}{rgb}{0.1,0.2,0.6}
\definecolor{darkred}{rgb}{0.8,0.1,0.2}
\definecolor{darkgreen}{rgb}{0.31,0.62,0.24}
\definecolor{bleudefrance}{rgb}{0.19, 0.55, 0.91}
\usepackage[colorlinks,citecolor=darkblue,linkcolor=darkblue,urlcolor=darkblue]{hyperref} 
\usepackage{enumerate}
\usepackage{setspace}
\usepackage{url}  % This makes \url work
\usepackage{mathrsfs}
\usepackage{mathtools} % for coloneqq

\newcommand{\hg}[1]{\textsc{h}_{#1}}

\newcommand{\cnot}[1]{\textsc{cnot}_{#1}}

\newcommand{\cz}[1]{\textsc{cz}_{#1}}

\begin{document}

\title{Quantum Volume for Photonic Quantum Processors}

\author{Yuxuan Zhang}
\affiliation{Cisco Quantum Lab, San Jose, CA 95134, USA}
\affiliation{Department of Physics, The University of Texas at Austin, Austin, TX 78712, USA}

\author{Daoheng Niu}
%\email{daoniu@cisco.com}
\affiliation{Cisco Quantum Lab, San Jose, CA 95134, USA}
\affiliation{Department of Physics, The University of Texas at Austin, Austin, TX 78712, USA}

\author{Alireza Shabani}
%\email{ashabani@cisco.com}
\affiliation{Cisco Quantum Lab, Los Angeles, CA 90049, USA}

\author{Hassan Shapourian}
\affiliation{Cisco Quantum Lab, San Jose, CA 95134, USA}

%\date{\today}

\begin{abstract}
Defining metrics for near-term quantum computing processors has been an integral part of the quantum hardware research and development efforts. Such quantitative characteristics are not only useful for reporting the progress and comparing different quantum platforms but also essential for identifying the bottlenecks and designing a technology roadmap. Most metrics such as randomized benchmarking and quantum volume were originally introduced for circuit-based quantum computers and were not immediately applicable to measurement-based quantum computing (MBQC) processors such as in photonic devices. In this paper, we close this long-standing gap by presenting a framework to map 
physical noises and imperfections in MBQC processes to logical errors in equivalent quantum circuits, whereby enabling the well-known metrics to characterize MBQC.
To showcase our framework, we study 
a continuous-variable cluster state based on the Gottesman-Kitaev-Preskill (GKP) encoding as a near-term candidate for photonic quantum computing, and derive the effective logical gate error channels and calculate the quantum volume in terms of the GKP squeezing and photon transmission rate.

\end{abstract}

\maketitle

\paragraph{Introduction}
Building a universal quantum computer beyond proof-of-principle demonstrations~\cite{wang2019boson,zhong2021phase,arute2019quantum} has been the holy grail of quantum computing research for years. Current mainstream superconducting circuit~\cite{arute2019quantum} and trapped-ion~\cite{pino2021demonstration} 
 platforms face challenges due to limitations in scalability~\cite{bruzewicz2019trapped} or lack of physical connectivity~\cite{kjaergaard2020superconducting}. On the other hand, there has been significant progress over the past decade in developing photonic-based platforms as quantum information processors~\cite{obrien2009photonic,bogdanov2017material}, which can provide some of the best solutions to these challenges: 
Photons are cost efficient to generate, and they are not physically restricted to a location and could be routed around. 
Moreover, quantum information encoded in photons, unlike inf atomic or solid-state systems, requires no quantum transduction for interchip communication, which can be crucial in distributed quantum computation~\cite{beals2013efficient}.

Admittedly, photonic computing does come with some drawbacks: It is generally hard to store photons unremittingly, and they can be lost during transmission and logical operations. The first drawback can be resolved by measurement-based quantum computing (MBQC) schemes~\cite{briegel2009measurement}, where one avoids the longevity problem by constantly teleporting quantum information from one photon to another one. One solution to the second issue is to encode quantum information in quadratures of optical fields, a.k.a.~continuous variable~\cite{lloyd1999quantum, menicucci2006universal}, as opposed to single photons.
Another route would be based on fault-tolerant encoding schemes~\cite{kitaev2003fault,bravyi1998quantum,mackay2004sparse,breuckmann2021quantum}, which allow one to address both the photon loss and Pauli errors at the cost of a large overhead. Although the latter attempts are impressive from a theoretical perspective, they all require revolutionary advancement in hardware manufacturing, such as faithfully preparing 3d cluster states~\cite{raussendorf2006a,raussendorf2007topological}.

In this work, we build a toolkit to address the following question: How does a measurement-based photonic quantum processor, to be developed in near future, perform compared to existing circuit-based quantum computing (CBQC) platforms? For this purpose, we need a metric. The quantum volume (QV)~\cite{cross2019validating} 
has been accepted as a common metric to measure the power of a noisy intermediate-scale quantum (NISQ) computer and to compare different circuit-based platforms.
QV has been routinely calculated and reported for CBQC platforms such as superconducting~\cite{ibm2022website} and trapped ion~\cite{baldwin2022re} technologies.
In this letter, we extend the QV formulation to measurement-based platforms enabling a common ground to compare the power of photonic quantum processors with other quantum processors.

\begin{figure}
    \centering
    \includegraphics[scale=0.5]{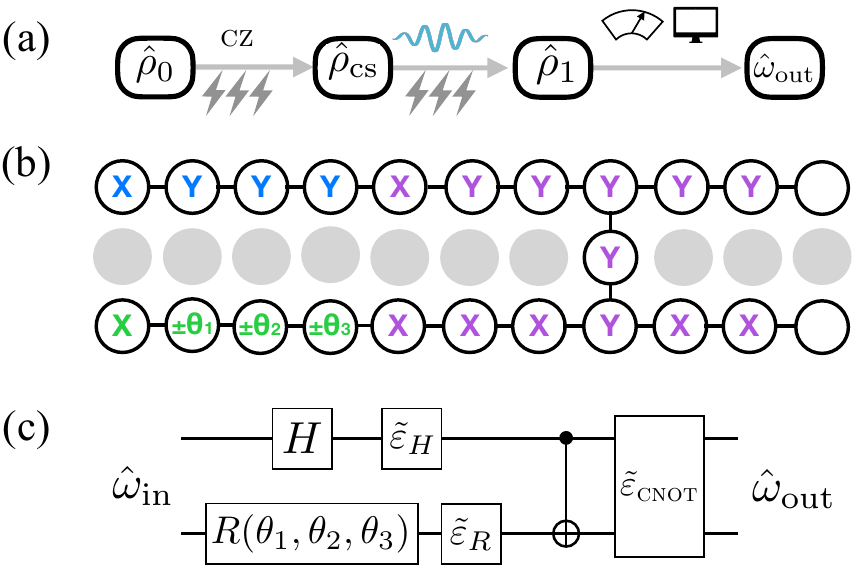}
    \caption{ {\bf Equivalent quantum circuit of a noisy measurement based quantum computation} \textbf{(a)} Diagram summarizing the noisy MBQC procedure. Noise is mainly generated during the cluster state preparation and measurements. \textbf{(b)} An instance of MBQC implementation of some logical gates. Each color represents a measurement pattern for one particular type of quantum gate, and letters denote specific measurement basis, following the convention of Raussendorf's original work ~\cite{raussendorf2003measurement}. \textbf{(c)} The corresponding noisy circuit. As we show in this paper, erroneous measurements on a noisy cluster state leads to an equivalent quantum circuit where each logical gate is accompanied by a Pauli channel (denoted by $\tilde{\varepsilon}$).
}
    \label{fig:demo}
\end{figure}

Our goal is to find a parameter regime for the photonic hardware as a universal quantum machine rather than a task-specific chip~\cite{mezher2022assessing,goel2022inverse,cavailles2022high} and at an intermediate scale rather than a fault-tolerant scale with millions of qubits~\cite{takeda2019toward,bourassa2021blueprint,bartolucci2021fusion,larsen2021fault}. %To this end, 
{As a direct simulation of the MBQC process is computationally demanding,} our main contribution is to show that implementing MBQC on noisy cluster states via faulty measurement devices is equivalent to perfect logical gates followed by a Pauli error channel (Fig.~\ref{fig:demo}). This error model is served as an effective description of photonic NISQ devices and makes them amenable to various NISQ benchmarks~\cite{magesan2011scalable,cross2019validating}. We derive this effective description in two cases: First, a discrete-variable (DV) cluster state as a warm-up example, where we consider depolarizing channels for imperfect resource state preparation and measurements. Second, a continuous-variable (CV) cluster state as an experimentally relevant setup~\cite{menicucci2006universal,pysher2011parallel,chen2014experimental,yokoyama2013ultra,roslund2014wavelength,wang2014weaving,yoshikawa2016invited,larsen2021deterministic} based on approximate Gottesman-Kitaev-Preskill (GKP) encoding~\cite{gottesman2001encoding}, where we include photon loss as the main source of error during cluster state generation and transmission. 

\paragraph{Framework}

The essence of the MBQC scheme is to use the cluster state and perform universal quantum computation with single-qubit measurements. 
Without making further assumptions on the physical platform, we  note  that both cluster state preparation and measurement steps can be noisy in realistic systems~\cite{usher2017noise}. This fact begs the following question: When a noisy quantum channel is applied to a cluster state $\hat \rho$ what happens to the measurement-based quantum operations performed on the encoded logical state $\hat \omega$?

As demonstrated in Fig.~\ref{fig:demo}{(a)}, in the initialization of MBQC %\DN{during the initialization of MBQC}, 
one is given $\hat{\rho}_0\coloneqq\hat{\omega}_{\text{in}}\otimes\ket{+}\bra{+}^{\otimes N_c}$ where $N_c$ is the number of \emph{channel} qubits to be measured. Without loss of generality, we assume $\hat{\rho}_0$ is a product state and can be prepared perfectly. The next step is to prepare the graph state $\hat \rho_{\text{cs}}$ in the desired form by applying a series of noisy controlled-phase gates which are modeled as perfect gates accompanied by a two-qubit depolarizing channel. 

\begin{figure}
    \centering
    \includegraphics[scale=0.4]{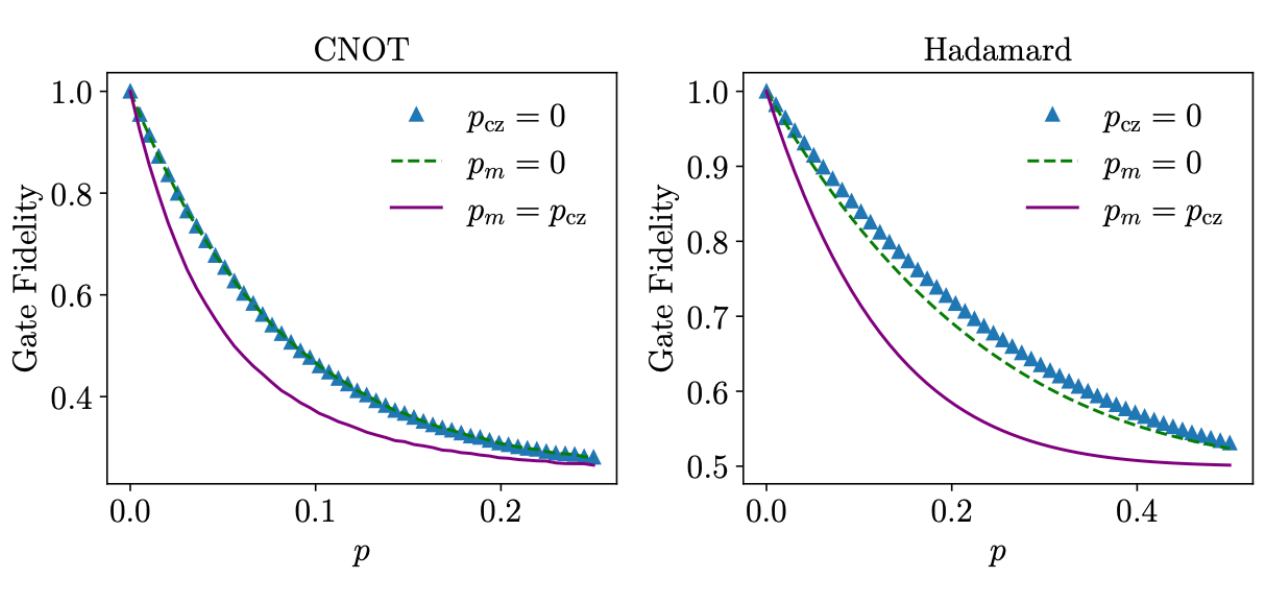}
    \caption{{\bf Logical gate fidelity} of two representative Clifford gates: Controlled-not and Hadamard against error rate $p$. We plotted three different cases: 1) $p_m = p,\  p_{\text{cz}}=0$; 2) $p_m = 0,\ p_{\text{cz}}=p$; 3) $p_m = p_{\text{cz}}=p$, where $p_m$ and $p_{\text{cz}}$ denote the error probability in single-qubit and two-qubit depolarizing channels, respectively. Notice that we consider the nearest neighbor $\cnot{}$ gate here, as non-nearest neighbor $\cnot{}$ gates can be realized with additional usage of optical swap gates, whose noise we are neglecting in this work.}
    \label{fig:dv-fidelity}
\end{figure}

Additionally, a faulty measurement apparatus can be modeled by a depolarizing channel followed by an ideal single-qubit projective measurement. 
{In an ideal measurement-based realization of a unitary gate $\hat U$, upon measuring the input and channel qubits we obtain an effective unitary operation $\hat P_{\Sigma_{\vec{s}}} \hat U$ (where $\hat P_{\Sigma_{\vec{s}}}$ is a Pauli operator~\cite{raussendorf2003measurement} depending on the measurement outcomes $\vec{s}=(s_1,\cdots,s_{N_c}), s_i=\pm$).} 
However, because of the Pauli errors defined above, the measurement results may be flipped.
This ultimately introduces an effective error channel {for each logical gate} which generally takes the form of a Pauli error channel 
 ${\tilde{\varepsilon}} (\hat{\omega})  = \sum_{\hat P\in \mathcal{P}_n} p_P \hat P \hat{\omega} \hat P$,
as shown for example in Fig.~\ref{fig:demo}(c), where $\mathcal{P}_n$ denotes the Pauli group on $n$ qubits and the error probabilities $p_P >0$ are derived for elementary gates in  Appendix~A~\cite{supp}. 
For instance, considering logical gates such as $\cnot{}$ or $\hg{}$, we find that the effective error channel are two-qubit and one-qubit Pauli error channels.
Figure~\ref{fig:dv-fidelity} shows the fidelity of these two gates at different noise rates. {Having the equivalent noisy circuit (such as the one shown in Fig.~\ref{fig:demo}(c)) at hand, we can run various benchmarks to simulate the performance of an MBQC scheme.}
 
The above derivation serves as a warm-up example since photons are lossy in realistic systems and any DV encoding suffers from a finite erasure probability even through concatenation with fault-tolerant schemes. Nonetheless, loss error in MBQC can be handled in CV encodings since loss does not lead to erasure and effectively degrades the qubit quality by reducing the squeezing level. 
In the following section, we investigate GKP encoding, where photon loss can be either corrected or transformed into a Pauli error, as a promising solution to performing MBQC in photonic systems.

%%%%%%%%%%%%%%%%%%%%%%%

\paragraph{GKP encoding}
This encoding is a quantum error correcting code based on the CV approach, in which qumodes, i.e., optical  (bosonic)  modes with two quadrature operators $\hat x = (\hat a^\dag+\hat a)/\sqrt{2}$ and $\hat p = i(\hat a^\dag-\hat a)/\sqrt{2}$, where $\hat a$ and $\hat a^\dag $ are annihilation and creation photon operators and $[\hat x, \hat p] = i$, are used to encode the quantum information (See Appendix B~\cite{supp} for a brief review of GKP qubits and details of various derivations discussed later in this section). The logical Hilbert space is two-dimensional (hence called GKP qubit), where ideal codewords  are defined by
\begin{align}
 \label{eq:idealGKP} 
 \ket{\mu_\text{gkp}} &= \sum_{n} \ket{(2n+\mu)\sqrt{\pi}}_x, \quad \mu=0,1.
\end{align}

An important property of GKP qubits is that the Clifford group can be implemented using only Gaussian operations. More explicitly, 
controlled-phase gate can be realized by Gaussian operations as in
$\cz{\text{gkp}}^{j, k} =  \exp[ - i \hat{x}_{j}\hat{x}_{k}]$.
Furthermore, to perform the Clifford gates in the MBQC scheme, we projectively measure GKP qubits in Pauli bases $(\hat X_\text{gkp},\hat Z_\text{gkp},\hat Y_\text{gkp})$ which can be implemented by homodyne measurements in one of the quadratures $\hat p$, $\hat x$, and the diagonal direction $\hat x-\hat p$, respectively.
On the other hand, realizing non-Clifford gates such as arbitrary single-qubit rotation in the measurement-based scheme requires measuring GKP qubits in non-Pauli bases, e.g.~along a Bloch vector characterized by the angle $\theta$ with respect to $\hat X_\text{gkp}$ measurement axis within the $\hat X_\text{gkp}\hat Y_\text{gkp}$ measurement plane as shown in Fig.~\ref{fig:demo}(a). A straightforward way to implement such non-Pauli measurement bases is by applying a unitary rotation around $\hat Z_\text{gkp}$ axis via $\hat U_{Z_\text{gkp}}(\theta) = \ket{0_\text{gkp}}\bra{0_\text{gkp}}+ e^{i\theta}\ket{1_\text{gkp}}\bra{1_\text{gkp}}$ (which can be implemented using methods in Refs.~\cite{gottesman2001encoding,yanagimoto2020engineering,konno2021non}), and then making a homodyne measurement of $\hat p$ quadrature. 

In what follows, we discuss the effective error models. Same as before, there are two instances where errors may happen: during the preparation and the measurement. 
Two types of errors may in turn occur during the cluster state preparation: The first is in generating GKP qubits, and the second is after applying each $\cz{}$ gate {(where the photon loss is modeled by a Lindblad master equation)}. Starting from a physical model for photonic gates, we find that faulty Clifford gates lead to Gaussian noise models whereas non-Clifford gate errors are non-Gaussian. We choose {an error ansatz for} the latter as a problem-agnostic depolarization channel on the physical level whose strength is related to the quality of input GKP qubits~\footnote{This assumption is partly motivated by the first-principle results of Refs.~\cite{yanagimoto2020engineering,konno2021non}.}.

\begin{figure*}
    \centering
    \includegraphics[scale=1]{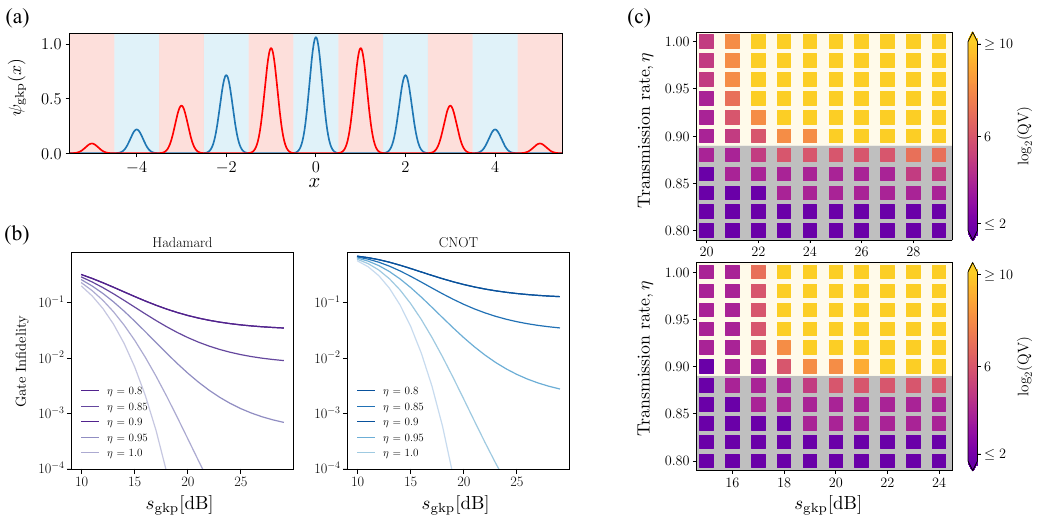}
\caption{{\bf Characteristics of MBQC on noisy GKP cluster states} {\bf(a)} A GKP wave function at squeezing rate $s_{\text{gkp}}= 12$ dB. As one decreases the squeezing rate, a red wave function will have more overlap in the blue region and vice versa, leading to a logical error. %; an ideal GKP requires infinite squeezing rate, which is unphysical.
The blue region is encoded as a $\ket{0_{\text{gkp}}}$ and red region as $\ket{1_{\text{gkp}}}$. {\bf(b)} Single gate infidelity versus squeezing for Hadamard (left) and $\cnot{}$ (right), and $\sigma_{\text{gkp}} = \sigma_{\text{cz}}$. {\bf(c)} Quantum volume plotted at various transmission rate and squeezing level, assuming $\sigma_{\text{gkp}} = \sigma_{\text{cz}}$ (top) and $\sigma_{\text{cz}} = 0$ (bottom), respectively. Grey area represents the transmission range that $\text{log}_2(\text{QV})< 10$ even when $s_{\text{gkp}}\rightarrow\infty$. Each point is averaged over 1,600 random instances to reduce the error on the heavy output probability to $\sim 1\%$.} 
    \label{fig:gkp}
\end{figure*}

The ideal GKP codewords in (\ref{eq:idealGKP}) are unphysical since they are given by an infinite sum (and hence non-normalizable) and cost infinite energy to construct. A realistic approximate GKP states \cite{menicucci2014fault} have a finite average photon number and can be modeled by  applying a Gaussian envelope operator to an ideal GKP state $\ket{\psi_\text{gkp}^\Delta} \propto e^{-\Delta \hat a^\dag \hat a} \ket{\psi_\text{gkp}}$.
In other words, an approximate GKP state can be understood as a coherent superposition of Gaussian peaks at ideal peak locations in Eq.~(\ref{eq:idealGKP}) along with a Gaussian envelope (see Fig.~\ref{fig:gkp}(a))~\cite{terhal2016encoding,shi2019fault,pantaleoni2020modular,tzitrin2020progress,matsuura2020equivalence}.
\begin{align}\label{eq:approximateGKP}
    \ket{\mu^\Delta_{\rm gkp}} \propto \int_{-\infty}^{+\infty} & ds\sum_{n}e^{-\Delta[(2\sqrt{\pi}+\mu)n]^2/2} \nonumber \\ &\times e^{-[s-(2\sqrt{\pi}+\mu)n]^2/2\Delta}\ket{s}_x.
\end{align}
The finite width of Gaussian peaks is usually referred to as finite squeezing level.
Analyzing the performance of such realistic GKP qubits is based on the twirling approximation, where a GKP state with finite squeezing is approximated by an incoherent mixture of random displacement errors as in  
\begin{align}
\ket{\psi_{\text{gkp} } } \rightarrow {\cal G}[\sigma](\ket{\psi_{\text{gkp} } }\bra{ \psi_{\text{gkp}} } ).
\label{eq:noisy-GKP-incoherent}
\end{align}
The error channel ${\cal G}[\sigma]$ randomly shifts the position and momentum quadratures by $\hat{x}\rightarrow \hat{x}+\xi_{x}$ and $\hat{p}\rightarrow \hat{p}+\xi_{p}$, where $\xi_{x}$ and $\xi_{p}$ follow a Gaussian random distribution with zero mean and standard deviation $\sigma$, i.e., $\xi_{x},\xi_{p}\sim {\cal N}(0,\sigma)$. 
More explicitly, the approximate GKP state (\ref{eq:approximateGKP}) is equivalent to a Gaussian noise with variance $\sigma_{\text{gkp}}^2=\Delta/2$  which is usually reported in dB as in $s_{\text{gkp}}\coloneqq -10\log_{10}(2\sigma_{\text{gkp}}^{2})$~\cite{menicucci2014fault,fukui2018high,fukui2019high}. 
Another feature of this approximation is that various other non-ideal effects during later stages of processing such as transmission loss and measurement inefficiency can be included by modifying the Gaussian noise variance~\footnote{We should note that the twirling approximation overestimates the error since we are dropping the coherent terms~\cite{noh2020fault}. Hence, this approximation while simplifying our analysis amounts to a more pessimistic error model and leads us to more conservative conclusions.}. 

The other source of error during the state preparation is caused by noisy $\cz{}$ gates and is modeled as an ideal $\cz{}$ followed by a correlated Gaussian random displacement error $\hat{x}_{i}\rightarrow \hat{x}_{i} + \xi_{x}^{(i)}$ and $\hat{p}_{i}\rightarrow \hat{p}_{i} + \xi_{p}^{(i)}$ for  $i\in \lbrace 1,2 \rbrace$, where the additive shift errors are drawn from bivariate Gaussian distributions $(\xi_{x}^{(1)},\xi_{p}^{(2)}) , (\xi_{x}^{(2)},\xi_{p}^{(1)}) \sim {\cal N}(0,\Sigma_{\text{cz}})$ with the noise covariance matrix 
proportional to $\sigma_{\text{cz}}^2 = \kappa /g$, where $\kappa$ and $g$ denote the loss rate and coupling coefficient, respectively (See Appendix B~\cite{supp} for derivation details).

Combining the Gaussian error channels associated with approximate GKP qubits and noisy $\cz{}$ gates, we obtain an expression for the noisy cluster state as in
\begin{align}
    \hat{\rho}_\text{cs} &= {\cal G}_{\boldsymbol{\Sigma}} \left(\hat U_{\text{cs}}\ket{+_{\text{gkp} } } \bra{ +_{\text{gkp}}  }^{\otimes N} \hat U_{\text{cs}}^\dag \right),
\end{align}
where $\hat U_{\text{cs}}$ is the ideal operator equal to a product of $\cz{}$ operators to generate the desired graph state.  
The above expression is a multimode version of (\ref{eq:noisy-GKP-incoherent}),  see Appendix B~\cite{supp} for details. 
As the final step of MBQC, we measure the GKP qubits on the cluster state.

{The photon loss during signal transmission and} inefficient homodyne measurements are modeled by passing the noisy GKP state through a fictional beam splitter before an ideal measurement is performed. Using the twirling approximation, the beam splitter can be modeled as a Gaussian random displacement error ${\cal N}[\sigma_m]$ of variance
$\sigma_m^2= (1-\eta)/2\eta$,
where {we call $0<\eta\leq 1$ the transmission rate.}
To implement the Clifford gates, we perform
 noisy homodyne measurements which are modeled by the transformation $\boldsymbol{\Sigma} \to \boldsymbol{\Sigma}+\sigma_m^2 \openone$ and an ideal measurement. Figure~\ref{fig:gkp}(b) shows the infidelity of measurement-based Clifford gates as a function of GKP squeezing $s_\text{gkp}$ for various values of the transmission rate $\eta$. An important observation here is that increasing the squeezing generally improves the gate fidelity as long as transmission rate  is large enough $\eta\geq 0.9$.

To implement an arbitrary logical qubit rotation, we need to make non-Pauli measurements, which can be realized by a non-Clifford single-qubit gate $\hat U_{Z_\text{gkp}}(\theta)$ before a homodyne measurement. Applying the noisy version of this unitary leads to a non-Gaussian error model which we model as a depolarizing error channel  where the error rate is determined by the fidelity of the state-of-the-art implementation of unitary operators~\cite{yanagimoto2020engineering,konno2021non} which in turn depends on the input state squeezing~\footnote{Notice that one has the freedom to change this general depolarizing channel to a scheme-specific one with a few changes in the derivation.}. As we show in Appendix~B~\cite{supp}, the effective error model of logical qubit rotations takes the form of a single-qubit Pauli error channel.

\paragraph{Quantum volume}
Here, we put {the noisy gates together (as shown e.g.~in Fig.~\ref{fig:demo}(c))} and use QV to examine the quality of multi-qubit MBQC carried out on a noisy GKP cluster state.
%.
The key idea of the QV analysis is the following question: Given a noisy quantum device, what is the largest $d\times d$ randomly permuted brick-wall circuit drawn from the Haar distribution it could execute with high fidelity? Exactly modeling the noisy channel executed by a quantum device requires gate-set tomography~\cite{nielsen2021gate}, which has an impractical, exponential sampling cost in system size. Nevertheless, one could instead use a heavy output string method ~\cite{aaronson2016complexity} to benchmark, which requires far fewer samples. The idea is based on the probability distribution $p_{U} (s)$ of output state of circuit $U$ as a bit string $s$ assuming that the input state is the all-zero state. The set of heavy outputs, $H_U$ is defined as
\begin{align*}
H_U = \{p_U(s), \ s \in \{0,1\}^m :\ p_U(s)>p_\text{med}\}.
\end{align*}
where $p_\text{med}$ is the median of $p_{U} (s)$. However, an actual probability distribution $q_U(s)$ generated by a noisy quantum computer may deviate from the ideal one, and the probability of obtaining a heavy output state {is denoted as} $h_{U} =\sum_{s\in H_U}{q(s)}$. An ideal evolution gives $h_{U} \sim 0.85$ whereas a random guess obviously gives $h_{U} = 0.5$. We follow the original convention in \cite{cross2019validating} and require $h_{U} \geq \frac{2}{3}$ for the condition of success, and define $QV\coloneqq 2^d$.

Figure~\ref{fig:gkp}(c) shows the QV benchmark results for GKP cluster states as a function of squeezing rate $s_{\text{gkp}}$ and transmission rate $\eta$ for two cases: {Whether $\cz{}$ gates to construct cluster state are lossy or not ($\sigma_{\cz{}}=0$)}. An immediate observation is that below $90\%$ transmission rate (shaded as gray regions), the QV is rather limited. This fact can also be seen from the gate fidelity plots (Fig.~\ref{fig:gkp}(b)) where the fidelity saturates regardless of $s_{\text{gkp}}$. Furthermore, reaching a QV of $2^{10}$, which is the state-of-the-art value in trapped ions~\cite{baldwin2022re}, requires 22~dB squeezing (18~dB if $\cz{}$ errors are ignored) considering 5\% {transmission loss}.
To put numbers in perspective, we note that the record value of squeezing in optical fields is $15$~dB~\cite{vahlbruch2016detection} (albeit for the vacuum state and not the GKP state) which is not far from the desired regime. Our analysis can also be relevant to other platforms such as
trapped-ion~\cite{fluhmann2019encoding,deneeve2022error} and superconducting circuits~\cite{campagne2020quantum}, where GKP qubits have been realized and the achieved squeezing is reportedly $7.5$-$9.5$~dB.
Overall, our result implies that experimental progress is required to reach the desired levels of squeezing.

\paragraph{Discussion}
In summary, we solved a long-standing open problem of addressing noise in MBQC schemes and laid the foundation for applying benchmarking tools to MBQC devices.
Starting from physical error models, we derive effective error models for measurement-based logical gates and show that standard metrics such as QV can be defined for the MBQC scheme. Our conclusions, based on the QV results regarding the required hardware quality (squeezing, etc.), 
might seem daunting at first glance, but can be improved in a few directions.
First, twirling approximation for noisy GKP qubits tends to overestimate the error and can be handled more carefully. Second, we believe with better MBQC circuit compilers (which reduce the overall quantum resource cost)~\cite{broadbent2009parallelizing}  or alternative forms of cluster states (which improve gate fidelity)~\cite{walshe2020continuous,walshe2021steamlined} the performance may significantly improve. Third, regardless of the compiler, performing non-Clifford measurements on GKP qubits with higher confidence is desirable. 
In this work, we used the CV MBQC scheme based on GKP encoding to deal with photon loss. It would be interesting to see if the DV MBQC scheme can be carried out in the presence of photon loss (or qubit erasure in this case).
Along the same line, it may be worth exploring the possibility of applying
 error mitigation techniques to near-term MBQC systems~\cite{temme2017error,endo2021hybrid,czarnik2021error,cerezo2021variational}, especially because  the effective noise channel is biased (see e.g., Fig.~1 in Appendix A~\cite{supp}).  Last but not least, our framework paved the way for efficiently simulating MBQC on noisy cluster states and opened new avenues for analyzing the effect of noise on the measurement-based implementations of various algorithms and quantum simulations~\cite{lee2022measurement}.

\paragraph{Acknowledgements}
We acknowledge insightful discussions with
Bing Qi, Stephen DiAdamo, Galan Moody, Yufei Ding, and Ramana Kompella.

\bibliography{refs.bib}

\clearpage
 \onecolumngrid


\section*{Supplementary material}

# Supplementary information for "Quantum Volume for Photonic Quantum Processors"

Yuxuan Zhang,[1,2] Daoheng Niu,[1,2] Alireza Shabani,[3] and Hassan Shapourian[1]

[1]*Cisco Quantum Lab, San Jose, CA 95134, USA*
[2]*Department of Physics, The University of Texas at Austin, Austin, TX 78712, USA*
[3]*Cisco Quantum Lab, Los Angeles, CA 90049, USA*

The supplementary information is organized as follows: In Appendix A, we discuss the derivation details of effective gate error channels in a general (discrete-variable) cluster state subject to depolarizing error channels. In particular, we show that an error-prone MBQC process on a noisy cluster state with faulty measurement devices can be viewed as a noisy quantum circuit where attributes such as the quantum volume and randomized benchmarking can be defined and used to quantify the performance. Appendix B is devoted to reviewing GKP qubits, explaining technical details of noise models, and deriving the effective logical gate fidelities. Table I below summarizes our notation throughout this paper.

| Type | Variable | Description |
|---|---|---|
| Error rates | $p$ | Raw physical Pauli error rates |
| | $p_{\text{err}}$ | Gaussian error function for GKP states |
| | $e$ | Measurement-flip error rate |
| | $q$ | Effective logical error rate |
| Density matrices | $\hat{\omega}_{\text{in}}/\hat{\omega}_{\text{out}}$ | Input/output logical density matrices |
| | $\hat{\rho}$ | Cluster state density matrix |
| Quantum channels | $\mathcal{E}$ | Noisy unitary channel on cluster states |
| | $\varepsilon$ | Pauli error channel on cluster states |
| | $\tilde{\mathcal{E}}$ | Noisy unitary channel on logical states |
| | $\tilde{\varepsilon}$ | Pauli error channel on logical states |
| | $\mathcal{M}$ | Projective measurement on cluster states |
| | $\mathcal{M}_e$ | Noisy projective measurement on noisy cluster states |
| | $\mathcal{G}$ | Gaussian noise channel |
| | $\mathcal{D}$ | Lindblad channel |
| | $\hat{D}$ | Displacement operator |

TABLE I. Summary of notations in the paper.

## CONTENTS

**Appendix A: Derivation of effective error models**

In this appendix, we derive effective error models for logical operations implemented on a noisy cluster state via the measurement-based scheme. Before we get into details, let us recap the three steps in implementing unitary operator $\hat{U}$ via the MBQC scheme:

1. *Initialization.* A cluster state can be defined on an arbitrary graph $G(V, E)$, by preparing each qubit (as a vertex on the graph $\forall v \in V$) in $|+\rangle$ state and applying a controlled-phase gate between $v_i, v_j$ for every pair of vertices connected via an edge, i.e., $\forall e(v_i, v_j) \in E$. If there is a desired input state $|\psi_{\text{in}}\rangle$, the starting vertices are replaced with $|\psi_{\text{in}}\rangle$ instead of $|+\rangle$.

2. *Measurement patterns corresponding to desired quantum gates.* Each unitary gate can be implemented with a specific sequence of single-qubit measurements, e.g., Fig. 1 of main text. Measurement outcomes are stored in classical memory.

3. *Classical post processing of measurement outcomes.* This is a standard procedure in teleportation based algorithms. On the output nodes, one gets $|\psi_{\text{out}}\rangle = \hat{P}_{\Sigma_{\vec{s}}} \hat{U} |\psi_{\text{in}}\rangle$ where $\hat{P}_{\Sigma_{\vec{s}}}$ denotes a Pauli operator which depends on measurement outcomes $\vec{s} = (s_1, \cdots, s_n)$ associated with $n$ measurements. Hence, measurement results of logical qubits need to be reinterpreted because of $\hat{P}_{\Sigma_{\vec{s}}}$.

We consider two sources of error: erroneous CZ gates and noisy measurement devices.

We model a noisy $\text{CZ}_{a,b}$ gate by an ideal gate accompanied by a fully depolarizing two-qubit error channel:

$$\varepsilon_{a,b}(\hat{\rho}) = (1 - p_{cz})\hat{\rho} + \frac{p_{cz}}{15} \sum_{\hat{P}_{a,b} \in \mathcal{P}_2 \setminus \{II\}} \hat{P}_{a,b} \hat{\rho} \hat{P}_{a,b}. \tag{A1}$$

The graph state we use to perform measurement-based computations mostly comprises of linear cluster states except for $T$ junctions necessary for CNOT gates. Here, we present the error model for a linear cluster state where CZ gates are applied sequentially. The effect of $T$ junctions in CNOT can be taken into account in a similar fashion (we included this effect in our gate fidelity plots in Fig. 2 of main text).

Consider a linear cluster state generated by $|\text{CS}_N\rangle = \bigotimes_n \text{CZ}_{n,n+1} |+\rangle^{\otimes N}$ where qubits are labeled by $n = 0, \cdots, N$ such that $\text{CZ}_{1,2}$ is applied first and so on. Similarly, projective measurements start from $n = 1$. The effective error channel after the preparation process due to noisy CZ gates (A1) is given by

$$\varepsilon_{n,n+1}(\hat{\rho}_{\text{cs}}) = (1 - \frac{14 p_c}{15})\hat{\rho}_{\text{cs}} + \frac{2 p_c}{15} \sum_{P \in E_1} \hat{P}_{n,n+1} \hat{\rho}_{\text{cs}} \hat{P}^\dagger_{n,n+1} + \frac{2 p_c}{15} \sum_{P' \in E_2} \hat{P}'_{n,n+1} \hat{\rho}_{\text{cs}} \hat{P}'^\dagger_{n,n+1}, \tag{A2}$$

where $\hat{\rho}_{\text{cs}} = |\text{CS}_N\rangle \langle \text{CS}_N|$ and

$$E_1 = \{\hat{X} \otimes \mathbb{1}, \hat{Y} \otimes \mathbb{1}, \hat{Z} \otimes \mathbb{1}, \mathbb{1} \otimes \hat{Z}\}, \tag{A3}$$

$$E_2 = \{\hat{X} \otimes \hat{Z}, \hat{Y} \otimes \hat{Z}, \hat{Z} \otimes \hat{Z}\}, \tag{A4}$$

corresponding to uncorrelated single-qubit and correlated two-qubit Pauli errors, respectively. It is important to note that correlated errors only happen between nearest neighbors. This is because the instantaneous state after each CZ gate is stabilizer state (i.e., a linear cluster) and errors can be simplified up to operators in the instantaneous stabilizer group.

We also include erroneous measurement devices which are modeled by a fully depolarizing channel

$$\varepsilon_n(\hat{\rho}) = (1 - p_m)\hat{\rho} + \frac{p_m}{3} \sum_{\hat{P}_n \in \mathcal{P}_1 \setminus \{I\}} \hat{P}_n \hat{\rho} \hat{P}_n, \tag{A5}$$

prior to projective measurements. Combining the two error channels, we obtain the full error model which may flip the actual measurement results. As we show in the next section, flipped measurements effectively lead to a noisy measure-based gates on logical state $\hat{\omega}$, i.e., an ideal measurement-based gate is accompanied by a Pauli error channel in the following form

$$\tilde{\varepsilon}_a(\hat{\omega}) = \sum_{\hat{P}_a \in \mathcal{P}_1} q_{P_a} \hat{P}_a \hat{\omega} \hat{P}_a, \tag{A6}$$

for single-qubit gates and

$$\tilde{\varepsilon}_{a,b}(\hat{\omega}) = \sum_{\hat{P}_{ab} \in \mathcal{P}_2} q_{P_{ab}} \hat{P}_{ab} \hat{\omega} \hat{P}_{ab}. \tag{A7}$$

for two-qubit gates (which in our case we only use measurement-based implementation of the CNOT gate). As we will see below, the effective Pauli error channel depends on the logical gate and is not generally isotropic, i.e., different Pauli errors come with different weights.

### 1. Effective gate error channels

#### a. Clifford gates

Consider a logical gate $\hat{U}$ which is implemented on a cluster state of $N = N_i + N_o$ qubits denoted by $\hat{\rho}$ where the measurement is performed on $j = 1, \cdots, N_i$ qubits and the result of logical operation is stored in qubits $j = N_i + 1, \cdots, N$. In what follows, we use the tensor product notation for operators as in $\hat{O}_i \otimes \hat{O}_o$ to keep track of the operator support over input and output qubits. The ideal quantum channel can be written as

$$\mathcal{M}(\hat{\rho}) = \sum_{\vec{s}} (\hat{\Pi}_{\vec{s}} \otimes \mathbb{1}_{N_o}) \hat{\rho} (\hat{\Pi}_{\vec{s}}^\dagger \otimes \mathbb{1}_{N_o}) \tag{A8}$$

where $\Pi_{\vec{s}}$ is the projection operator

$$\hat{\Pi}_{\vec{s}} = \bigotimes_{j=1}^{N_i} \left( \frac{1 + (-1)^{s_j} \hat{P}_j}{2} \right), \tag{A9}$$

$\vec{s} = (s_1, \cdots, s_{N_i}) \in \mathbb{Z}_2^{N_i}$ denotes the vector containing the measurement outcomes (i.e., an array of zeros and ones) and $\hat{P}_j \in \mathcal{P}_1$ is a single-qubit Pauli operator acting on qubit $j$ determined by the measurement basis. Measurement-based implementation of $\hat{U}$ then means to have

$$(\hat{\Pi}_{\vec{s}} \otimes \mathbb{1}_{N_o}) \hat{\rho}_{\mathrm{cs}} (\hat{\Pi}_{\vec{s}}^\dagger \otimes \mathbb{1}_{N_o}) = \hat{\Pi}_{\vec{s}} \otimes \left( \hat{P}_{\Sigma_{\vec{s}}} \hat{U} \left| \Psi_{\mathrm{in}} \right\rangle \left\langle \Psi_{\mathrm{in}} \right| \hat{U}^\dagger \hat{P}_{\Sigma_{\vec{s}}}^\dagger \right) \tag{A10}$$

for a sequence of measurement outcome $\vec{s}$, where $\hat{P}_{\Sigma_{\vec{s}}}$ is a Pauli operator acting on the output qubits. Now, suppose the cluster state is subject to a quantum error channel $\varepsilon$ before the measurement, i.e., $\mathcal{M}_e(\hat{\rho}) = \mathcal{M}(\varepsilon(\hat{\rho}))$. As we show below, this process can be viewed as an ideal measurement accompanied by a noisy channel; in other words, a measurment-based scheme on a noisy cluster state realizes a noisy logical operator.

$$\begin{aligned}
\mathcal{M}_e(\hat{\rho}) &= \sum_{\vec{s}} (\hat{\Pi}_{\vec{s}} \otimes \mathbb{1}_{N_o}) \varepsilon(\hat{\rho}) (\hat{\Pi}_{\vec{s}}^\dagger \otimes \mathbb{1}_{N_o}) \\
&= \sum_{\vec{s}} \sum_{\mu} p_\mu (\hat{\Pi}_{\vec{s}} \otimes \mathbb{1}_{N_o}) \hat{K}_\mu \hat{\rho} \hat{K}_\mu^\dagger (\hat{\Pi}_{\vec{s}}^\dagger \otimes \mathbb{1}_{N_o}) \\
&= \sum_{\mu} p_\mu \sum_{\vec{s}} \hat{K}_\mu (\hat{\Pi}_{\vec{s}_\mu} \otimes \mathbb{1}_{N_o}) \hat{\rho} (\hat{\Pi}_{\vec{s}_\mu}^\dagger \otimes \mathbb{1}_{N_o}) \hat{K}_\mu^\dagger \\
&= \sum_{\mu} p_\mu \hat{K}_\mu \left( \sum_{\vec{s}} (\hat{\Pi}_{\vec{s}} \otimes \mathbb{1}_{N_o}) \hat{\rho} (\hat{\Pi}_{\vec{s}}^\dagger \otimes \mathbb{1}_{N_o}) \right) K_\mu^\dagger \\
&= \sum_{\mu} p_\mu \hat{K}_\mu \mathcal{M}(\hat{\rho}) \hat{K}_\mu^\dagger.
\end{aligned} \tag{A11}$$

Here, $\varepsilon(\hat{\rho}) = \sum_\mu p_\mu \hat{K}_\mu \hat{\rho} \hat{K}_\mu^\dagger$ where $\hat{K}_\mu = \hat{K}_{\mu,i} \otimes \hat{K}_{\mu,o}$ is a Pauli operator which may act on both input and output qubits. Swapping the opertors $\hat{\Pi}_{\vec{s}} \hat{K}_{\mu,i}$ shifts the $\vec{s}$ as in $\vec{s}_\mu = \vec{s} + \vec{r}_\mu$ in which $\vec{r}_\mu$ is a $\mathbb{Z}_2$ vector with non-zero entries wherever $\hat{P}_j$ (see the definition of $\hat{\Pi}_{\vec{s}}$ in Eq. (A9)) and $\hat{K}_\mu$ anticommute. To go from the third line to the fourth line, we relabel the summation variable $\vec{s}$. Plugging in Eq. (A10) to Eq. (A11), we find that the noise channel on input

qubits leads to a noise channel on output qubits:

$$\begin{aligned}
\mathcal{M}_e(\hat{\rho}_{\text{cs}}) &= \sum_\mu p_\mu \hat{K}_\mu \left( \sum_{\vec{s}} \hat{\Pi}_{\vec{s}} \otimes \left( \hat{P}_{\Sigma_{\vec{s}}} \hat{U} |\Psi_{\text{in}}\rangle \langle \Psi_{\text{in}}| \hat{U}^\dagger \hat{P}^\dagger_{\Sigma_{\vec{s}}} \right) \right) \hat{K}^\dagger_\mu \\
&= \sum_\mu p_\mu \sum_{\vec{s}} (\hat{K}_{\mu,i} \hat{\Pi}_{\vec{s}} \hat{K}^\dagger_{\mu,i}) \otimes \left( \hat{K}_{\mu,o} \hat{P}_{\Sigma_{\vec{s}}} \hat{U} |\Psi_{\text{in}}\rangle \langle \Psi_{\text{in}}| \hat{U}^\dagger \hat{P}^\dagger_{\Sigma_{\vec{s}}} \hat{K}^\dagger_{\mu,o} \right) \\
&= \sum_\mu p_\mu \sum_{\vec{s}} \hat{\Pi}_{\vec{s}_\mu} \otimes \left( \hat{K}_{\mu,o} \hat{P}_{\Sigma_{\vec{s}}} \hat{U} |\Psi_{\text{in}}\rangle \langle \Psi_{\text{in}}| \hat{U}^\dagger \hat{P}^\dagger_{\Sigma_{\vec{s}}} \hat{K}^\dagger_{\mu,o} \right) \\
&= \sum_\mu p_\mu \sum_{\vec{s}} \hat{\Pi}_{\vec{s}} \otimes \left( \hat{K}_{\mu,o} \hat{P}_{\Sigma_{\vec{s}_\mu}} \hat{U} |\Psi_{\text{in}}\rangle \langle \Psi_{\text{in}}| \hat{U}^\dagger \hat{P}^\dagger_{\Sigma_{\vec{s}_\mu}} \hat{K}^\dagger_{\mu,o} \right) \\
&= \sum_\mu p_\mu \sum_{\vec{s}} \hat{\Pi}_{\vec{s}} \otimes \left[ \left( \hat{K}_{\mu,o} \hat{P}_{\Sigma_{\vec{r}_\mu}} \right) \left( \hat{P}_{\Sigma_{\vec{s}}} \hat{U} |\Psi_{\text{in}}\rangle \langle \Psi_{\text{in}}| \hat{U}^\dagger \hat{P}^\dagger_{\Sigma_{\vec{s}}} \right) \left( \hat{K}_{\mu,o} \hat{P}_{\Sigma_{\vec{r}_\mu}} \right)^\dagger \right] \\
&= \sum_\mu p_\mu (\mathbb{1}_{N_i} \otimes \hat{K}_{\mu,o} \hat{P}_{\Sigma_{\vec{r}_\mu}}) \left( \sum_{\vec{s}} \hat{\Pi}_{\vec{s}} \otimes \left( \hat{P}_{\Sigma_{\vec{s}}} \hat{U} |\Psi_{\text{in}}\rangle \langle \Psi_{\text{in}}| \hat{U}^\dagger \hat{P}^\dagger_{\Sigma_{\vec{s}}} \right) \right) (\mathbb{1}_{N_i} \otimes \hat{K}_{\mu,o} \hat{P}_{\Sigma_{\vec{r}_\mu}})^\dagger
\end{aligned}$$

which can be recast as $\tilde{\varepsilon}(\mathcal{M}(\hat{\rho}_{\text{cs}}))$. Hence, the noise channel on logical (output) state can be written in a more compact form

$$\tilde{\varepsilon}(\hat{\omega}) = \sum_\mu p_\mu (\hat{K}_{\mu,o} \hat{P}_{\Sigma_{\vec{r}_\mu}}) \hat{\omega} (\hat{P}_{\Sigma_{\vec{r}_\mu}} \hat{K}_{\mu,o})^\dagger. \tag{A12}$$

Physically, what happens is that the measurement outcome $\vec{s}$ is flipped to $\vec{s}_\mu$ by the noise operator. As a result, compared to the ideal case (A10) there is an extra Pauli operator $\hat{P}_{\Sigma_{\vec{r}_\mu}}$ which depends on the noise operator $\hat{K}_\mu$. Having found the noise model, we can define an average fidelity for a measurement-based implementation $\mathcal{M}_e$ of logical operator $\hat{U}$ using the following expression [1],

$$\overline{\mathcal{F}}(\mathcal{M}_e, \hat{U}) = \frac{\sum_j \text{tr}\left( \hat{U} \hat{U}^\dagger_j \hat{U}^\dagger \tilde{\mathcal{E}}(\hat{U}_j) \right) + d^2}{d^2(d+1)}, \tag{A13}$$

where $d = 2^N$ is the Hilbert space dimension, and $\{\hat{U}_j\}$ is a set of Pauli operators which form an orthogonal basis of unitary operators such that $\text{tr}(U^\dagger_j U_k) = \delta_{jk} d$. For a single-qubit logical gate, we can write

$$\overline{\mathcal{F}}(\mathcal{M}_e, \hat{U}) = \frac{1}{2} + \frac{1}{12} \sum_{j=1,2,3} \text{tr}\left( \hat{U} \hat{\sigma}_j \hat{U}^\dagger \tilde{\mathcal{E}}(\hat{\sigma}_j) \right), \tag{A14}$$

where $\sigma_j \in \mathcal{P}_1$. The average fidelity of the single-qubit Clifford gates can further be simplified into

$$\overline{\mathcal{F}}(\mathcal{M}_e, \hat{U}_{\text{Cliff}}) = 1 - \frac{2}{3}(q_x + q_y + q_z), \tag{A15}$$

where $q_i$ is the logical Pauli error probability defined in Eq. (A6).

In the remainder of this section, we explicitly derive the effective noise channel for the several elementary gates which are the building blocks of the MBQC.

1. Identity gate:

    This gate is implemented on by a measuring the first two qubits of a 3-qubit cluster state in $X$-basis, and the state will be teleported from site 1 to 3, i.e., $\hat{P}_1 = \hat{P}_2 = \hat{X}$ in Eq. (A9). The measurement-outcome dependent Pauli operator is given by

    $$\hat{P}_{\Sigma_{\vec{s}}} = \hat{X}^{s_2} \hat{Z}^{s_1}, \tag{A16}$$

    where $\vec{s} = (s_1, s_2)$ denotes the measurement outcome of the first two qubits on the linear cluster. Considering

measurement-device depolarizing channel (A5), the effective error channel is given by (A6) where

$$q_x = q_z = \frac{2}{3}p_m(1 - \frac{2}{3}p_m),$$
$$q_y = \frac{4}{9}p_m^2. \tag{A17}$$

The effect of noisy cluster state (A2) can be similarly included; however, the derivation is rather tedious and we do not present the final full expression here. To the linear order in $p_m$ and $p_{\text{CZ}}$ the error probabilities are given by

$$q_x = q_z = \frac{2}{3}p_m + \frac{8}{15}p_{\text{CZ}},$$
$$q_y = \frac{8}{15}p_{\text{CZ}}. \tag{A18}$$

2. Hadmard gate:

Hadamard gate is realized by the measurement sequence $\hat{X}_1\hat{Y}_2\hat{Y}_3\hat{Y}_4$ on a 5-qubit cluster state [2], where we get

$$\hat{P}_{\Sigma_{\vec{s}}} = \hat{X}^{s_1+s_3+s_4}\hat{Z}^{s_2+s_3}, \tag{A19}$$

at the output qubit 5. Here, $\vec{s}$ is a four-component vector. The result of each single-qubit measurement can be flipped result by probability $e_m = 2p_m/3$ according to the depolarizing channel (A5). For instance, measuring $\hat{X}_1$ may be preceded by $\hat{Y}$ or $\hat{Z}$ error which flips the outcome. Assuming all the measurement errors are independent processes, we arrive at the Pauli error channel (A6) where

$$q_x = 2e_m(1-e_m)^3 + e_m^2(1-e_m)^2 + e_m^4,$$
$$q_z = q_y = e_m(1-e_m)^3 + 2e_m^2(1-e_m)^2 + e_m^3(1-e_m). \tag{A20}$$

Including the preparation error (A2), the error rates to linear linear are found to be

$$q_x = \frac{4}{3}p_m + \frac{16}{15}p_{\text{CZ}},$$
$$q_z = q_y = \frac{2}{3}p_m + \frac{16}{15}p_{\text{CZ}}. \tag{A21}$$

The average gate fidelity (A13) using the exact formulas for the error rates is plotted in Fig. 2 of main text. for three cases: when both error channels are present with $p_m = p_{\text{CZ}}$ and when only one channel is present.

3. Phase gate:

Phase gate is realized by the measurement sequence $\hat{X}_1\hat{X}_2\hat{Y}_3\hat{X}_4$ on a 5-qubit cluster state [2], which leads to a measurement-dependent auxiliary Pauli operator

$$\hat{P}_{\Sigma_{\vec{s}}} = \hat{X}^{s_2+s_4}\hat{Z}^{s_1+s_2+s_3}, \tag{A22}$$

at the output qubit 5, where $\vec{s}$ is a four-component vector. Similar to the Hadamard gate, the effective error model when only faulty measurement device is considered is given by Eq. (A6) where the error rates read

$$q_z = 2e_m(1-e_m)^3 + e_m^2(1-e_m)^2 + e_m^4,$$
$$q_x = q_y = e_m(1-e_m)^3 + 2e_m^2(1-e_m)^2 + e_m^3(1-e_m). \tag{A23}$$

Note that this is only different from the Hadamard for $q_x \leftrightarrow q_z$. Furthermore, the error rates in the presence of CZ-gate errors are modified as in

$$q_x = q_y = \frac{2}{3}p_m + \frac{16}{15}p_{\text{CZ}},$$
$$q_z = \frac{4}{3}p_m + \frac{16}{15}p_{\text{CZ}}, \tag{A24}$$

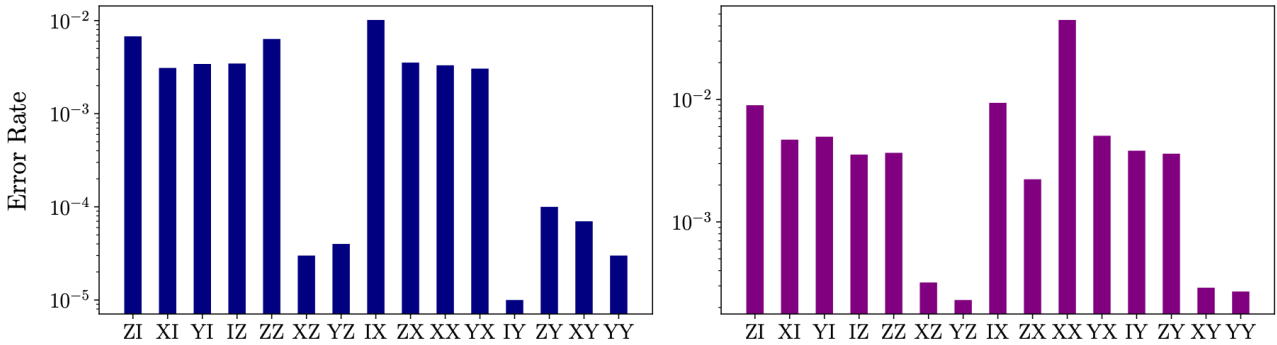

FIG. 1. **Effective error channels of the nearest neighbor cnot gate. Left:** $p_{\text{CZ}} = 0$, $p_m = 0.05$, **Right:** $p_{\text{CZ}} = p_m = 0.05$.

up to linear order in error probabilities.

4. CNOT gate:

We use the following scheme [2],

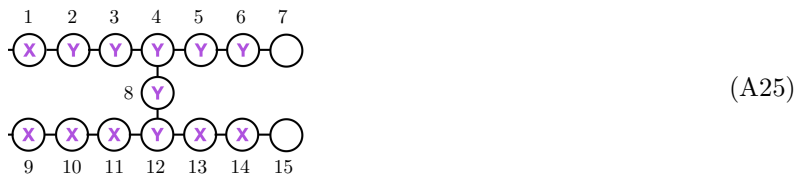
(A25)

to implement a CNOT$_{c,t}$ gate. The measurement-dependent Pauli operator can be written as

$$\hat{P}_{\Sigma_{\vec{s}}} = \hat{X}_c^{\gamma_x^{(c)}} \hat{X}_t^{\gamma_x^{(t)}} \hat{Z}_c^{\gamma_z^{(c)}} \hat{Z}_t^{\gamma_z^{(t)}}, \tag{A26}$$

where

$$\begin{aligned}
\gamma_x^{(c)} &= s_2 + s_3 + s_5 + s_6, \\
\gamma_x^{(t)} &= s_2 + s_3 + s_8 + s_{10} + s_{12} + s_{14}, \\
\gamma_z^{(c)} &= s_1 + s_3 + s_4 + s_5 + s_8 + s_9 + s_{11}, \\
\gamma_z^{(t)} &= s_9 + s_{11} + s_{13}.
\end{aligned} \tag{A27}$$

The effective error model reads as Eq. (A7), where we need to calculate 15 coefficients. We do not show the explicit expressions here as they are too long. Instead, the error rates as a function of depolarizing channel probability $p_m$ is shown in Fig. 1. It is interesting to note that different Pauli errors appear with different rates despite the fact that the original depolarizing channel is fully isotropic. Also, for large $p$, the error channel converges to the identity channel, $\tilde{\varepsilon}(\hat{\omega}) \to \text{tr}_{c,t}(\hat{\omega}) \otimes \mathbb{1}$. We further show the effective gate fidelity in Fig. 2 of main text for the three cases with and without the measurement-device and controlled-phase gate errors. It is worth noting that the fidelity in the two cases where we consider only one error channel behave quite similarly.

b. *Non-Clifford gates*

To achieve a universal gate set, we use arbitrary single-qubit rotation as a non-Clifford gate. A single qubit gate corresponding to an arbitrary rotation on the Bloch sphere can be decomposed in three rotations along x-z-x axes [2]. That is:

$$\hat{U}(\theta_1, \theta_2, \theta_3) = \text{RX}(\theta_3)\text{RZ}(\theta_2)\text{RX}(\theta_1). \tag{A28}$$

Equivalently, this can be accomplished by adaptive measurements on four qubits in a linear cluster state, (as shown in Fig. 1(a) of main text), where besides the first qubit which is measured in $\hat{X}$ basis, the other three qubits along the Bloch vector $\vec{r}_i = (\cos\theta_i, \sin\theta_i, 0)$ in $xy$-plane with angles $(-1)^{s_1+1}\theta_1$, $(-1)^{s_2+1}\theta_2$, and $(-1)^{s_1+s_3+1}\theta_3$, respectively. The overall correction Pauli operator is given by

$$\hat{P}_{\Sigma\vec{s}} = \hat{X}^{s_2+s_4}\hat{Z}^{s_1+s_3}. \tag{A29}$$

Analytically deriving the effective noise model for rotation gates is more involved, since the angles $\theta_i$ depends on previously measured results, i.e., $\pm\theta_i$. As we show below, however, the effective gate fidelity does not depend on the rotation angle when considering the native error probabilities to the first order. This means that the logical Pauli error channel is not gate dependent at this order of approximation, which is an important observation and simplifies the simulations significantly. We explicitly show this property for one of the non-Pauli basis measurements below and provide numerical evidence for angle-independence fidelity in a two-dimensional parameter space $(\theta_2, \theta_3)$ in Fig. 2.

Let us consider the RZ($\theta$) logical gate which is implemented by measuring four qubits on a five-qubit cluster state where qubits 1, 2, 4 are measured in the $\hat{X}$ basis and qubit 3 is measured at angle $(-1)^{s_2+1}\theta$ as explained above. We distinguish two scenarios, $\mathcal{M}_e(\hat{\omega}) = \mathcal{M}_{\text{noflip}}(\hat{\omega}) + \mathcal{M}_{\text{flipped}}(\hat{\omega})$, based on the correctness of $s_2$ [3]:

1. If no error occurs on qubit 2, then the Clifford gate error model derived in previous section applies. Since we are assuming an isotropic error channel, the measurement-flip error of the third qubit is independent of the measurement basis, though a more careful analysis is required for biased error models. Thus, the measurement-based logical operator is given by

$$\mathcal{M}_{\text{noflip}}(\hat{\omega}) = q_0 \text{RZ}(\theta)\hat{\omega}\text{RZ}(-\theta) + q_x\hat{X}\text{RZ}(\theta)\hat{\omega}\text{RZ}(-\theta)\hat{X} + q_y\hat{Y}\text{RZ}(\theta)\rho\text{RZ}(-\theta)\hat{Y} + q_z\hat{Z}\text{RZ}(\theta)\hat{\omega}\text{RZ}(-\theta)\hat{Z}, \tag{A30}$$

where we use the fact that $\text{RZ}(\theta)^\dagger = \text{RZ}(-\theta)$.

2. Measurement of qubit 2 is flipped by the noise. This way, we are not implementing RZ($\theta$) but RZ($-\theta$). Hence, the logical operator can be written as

$$\mathcal{M}_{\text{flipped}}(\hat{\omega}) = q'_0\text{RZ}(-\theta)\hat{\omega}\text{RZ}(\theta) + q'_x\hat{X}\text{RZ}(-\theta)\hat{\omega}\text{RZ}(\theta)\hat{X} + q'_y\hat{Y}\text{RZ}(-\theta)\hat{\omega}\text{RZ}(\theta)\hat{Y} + q'_z\hat{Z}\text{RZ}(-\theta)\hat{\omega}\text{RZ}(\theta)\hat{Z}. \tag{A31}$$

Next, we plug in the above expression to Eq. (A14) and obtain

$$\overline{\mathcal{F}}(\mathcal{M}_{\text{flipped}}, \text{RZ}) = \frac{1}{2} + \frac{1}{6}((1+2\cos(2\theta))q'_0 - q'_x - q'_y + (1-2\cos(2\theta)q'_z). \tag{A32}$$

Next, we use Eq. (A29) to derive the $q_i$ coefficients. Assuming no-flip on second qubit case, the following error configurations correspond to a Pauli error (1 stands for a flip) $\{I : 0000, 1010; X : 0001; 1011; Y : 1001, 1010; Z : 1000, 0010\}$.

$$\begin{aligned} q_x &= e_m(1-e_m)^3 + e_m^3(1-e_m), \\ q_y &= 2e_m^2(1-e_m)^2, \\ q_z &= 2e_m(1-e_m)^3, \\ q_0 &= (1-e_m)^4 + e_m^2(1-e_m)^2. \end{aligned} \tag{A33}$$

In the case of an erroneous second qubit, the following error configurations correspond to a Pauli error, $\{I : 0101, 1111; X : 0100; 1110; Y : 1100, 0110; Z : 1101, 0111\}$, from where we obtain

$$\begin{aligned} q'_x &= e_m(1-e_m)^3 + e_m^3(1-e_m), \\ q'_y &= 2e_m^2(1-e_m)^2, \\ q'_z &= 2e_m^3(1-e_m), \\ q'_0 &= e_m^4 + e_m^2(1-e_m)^2. \end{aligned} \tag{A34}$$

Putting everything together, $\overline{\mathcal{F}}(\mathcal{M}_e, \text{RZ}) = \overline{\mathcal{F}}(\mathcal{M}_{\text{noflip}}, \text{RZ}) + \overline{\mathcal{F}}(\mathcal{M}_{\text{flipped}}, \text{RZ})$; here, we have obtained an error model for single axis rotations (it also applies to other rotation gates due to symmetry in error model). It is not hard to check that setting $\theta = n\pi/2$ returns the same logical error rate as that of the Clifford gates $H$ and $S$. The $\theta$-dependent correction is higher order in $e_m$, and thus will not significantly affect the result, as shown in Fig. 2. The formula for general single qubit rotation, however, would be too tedious to derive. Thus, we use a numerical simulation instead to

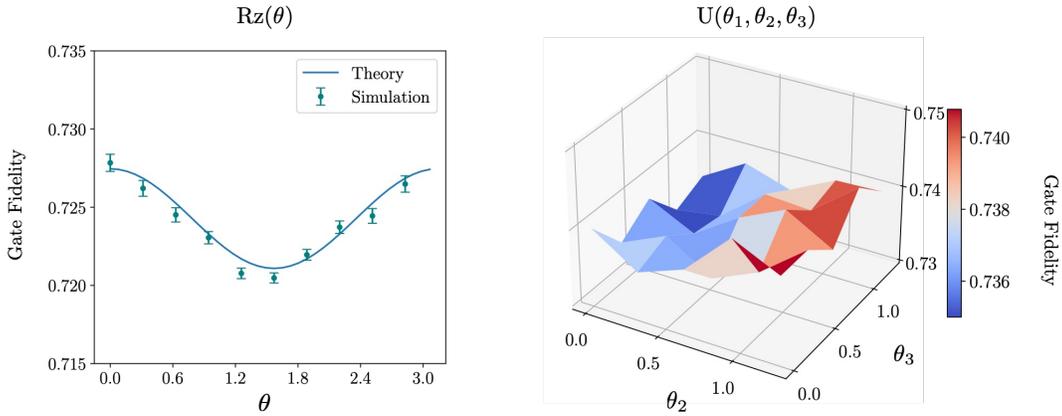

FIG. 2. **Average gate fidelity of rotation gates** We study the angle dependence of the rotation gates' average fidelity. **Left:** Theoretical and simulation results for RZ gate at $p_{\text{CZ}} = 0$, $p_m = 0.2$. **Right:** Qiskit simulation results for arbitrary rotation gate U at $p_{\text{CZ}} = 0$, $p_m = 0.15$; we fix $\theta_1 = \pi/4$. Overall, we find that the average performance of a rotation gate is relatively independent of its rotation angles.

show that the angle-dependent correction is small (see Fig. 2). The important conclusion here is that our treatment of effective error model for Clifford gates in the previous section is also applicable to arbitrary rotation gates (which are necessary for non-Clifford operations) up to lowest order in error probabilities.

### 2. Simulating MBQC on noisy cluster states

We explicitly derived error models for measurement-based gates in the previous sections. In what follows, we show that these logical gate error channels can be applied to the case of a quantum circuit consisting of multiple gates and hence generalizes to any measurement-based computation task. Consider circuit $U$ with a set of gates $g_1, g_2, \cdots, g_N$. The ideal MBQC circuit yields the following overall operator

$$|\psi_{\text{out}}\rangle = \left(\prod_{i=1}^{N} \hat{P}_{\Sigma, g_i}\Big|_U\right) \left(\prod_{i=1}^{N} \hat{U}_{g_i}\right) |\psi_{\text{in}}\rangle, \tag{A35}$$

which is built out of a series of measurement-based gates [2] as in

$$\left(\prod_{i=1}^{N} \hat{P}_{\Sigma, g_i}\Big|_U\right) \left(\prod_{i=1}^{N} \hat{U}_{g_i}\right) = \prod_{i=1}^{N} \left(\hat{P}_{\Sigma, g_i} \hat{U}_{g'_i}\right). \tag{A36}$$

Here, the net Pauli operator at the final output after propagating all $\hat{P}_{\Sigma, g_i}$ operators (associated with measurement outcomes implementing $g_i$) through the next set of $U_{g'_i}$ becomes $\prod_{i=1}^{N} P_{\Sigma, g_i}|_U$. We should also note that propagating each Pauli operator $\hat{P}_{\Sigma, g_i}$ through subsequent gates $\hat{U}_{g'_k}$, $k > i$ may change the unitary operator if it is a non-Clifford gate. In other words, we have

$$\hat{U}_{g'_k} = \begin{cases} \hat{U}_{g_k}, & \text{if } g_k \in \text{Cliff}, \\ \hat{P}_{\Sigma_k} \hat{U}_{g_k} \hat{P}_{\Sigma_k}, & \text{Otherwise.} \end{cases} \tag{A37}$$

where $\hat{P}_{\Sigma_k} = \prod_{i<k} \hat{P}_{\Sigma, g_i}|_{U_k}$ is the accumulated Pauli operator after propagating through the circuit up to depth $k$.

Let us now discuss how errors can be simulated. Consider a Pauli error $\hat{E}_{g_i}$ as a result of the noisy implementation of gate $g_i$; thus, we have

$$\prod_{i=1}^{N} \left(\hat{E}_{g_i} \hat{P}_{\Sigma, g_i} \hat{U}_{g'_i}\right) = \pm \left(\prod_{i=1}^{N} \hat{P}_{\Sigma, g_i}\right) \left(\prod_{i=1}^{N} \hat{E}_{g_i} \hat{U}_{g_i}\right), \tag{A38}$$

where we propagate $\hat{P}_{\Sigma,g_i}$ through both types of operators: the actual unitary $\hat{U}_{g'_i}$ same as before and the Pauli error $\hat{E}_{g_i}$. As we see there is not much change other than the sign factor, which is a cumulative sign based on the commutation relations of the $\hat{P}_{\Sigma,g_i}$'s and $\hat{E}_{g_j}$'s to be swapped. Notice that the sign factor does not matter after all, since the error channel is in terms of density matrices (see e.g., Eq. (A12)). Therefore, noisy MBQC is equivalent to having a noisy quantum circuit and can be simulated by a series of noisy gates, where each gate may be accompanied by a Pauli error as in $\hat{E}_{g_i}\hat{U}_{g_i}$.

### Appendix B: Error model for GKP cluster states

In this appendix, we give a brief review of some basics of GKP encoding in bosonic modes and provide details of the noise models (See also recent reviews such as Ref. [4]).

#### 1. Review of GKP states

As mentioned in the main text, the GKP qubit is a quantum error correcting code [5], where the ideal codewords on square lattice are defined by

$$|0_{\text{gkp}}\rangle = \sum_{n \in \mathbb{Z}} |2n\sqrt{\pi}\rangle_x, \qquad |1_{\text{gkp}}\rangle = \sum_{n \in \mathbb{Z}} |(2n+1)\sqrt{\pi}\rangle_x, \tag{B1}$$

Also, the complementary basis states $|\pm_{\text{gkp}}\rangle := \frac{1}{\sqrt{2}}(|0_{\text{gkp}}\rangle \pm |1_{\text{gkp}}\rangle)$ are given by

$$|+_{\text{gkp}}\rangle = \sum_{n \in \mathbb{Z}} |2n\sqrt{\pi}\rangle_p, \qquad |-_{\text{gkp}}\rangle = \sum_{n \in \mathbb{Z}} |(2n+1)\sqrt{\pi}\rangle_p. \tag{B2}$$

The GKP states form a quantum error correcting code in the 2-dimensional subspace of a single bosonic mode Hilbert space stabilized by the two generators

$$\hat{S}_x := \exp[i2\sqrt{\pi}\hat{x}], \qquad \hat{S}_p := \exp[-i2\sqrt{\pi}\hat{p}]. \tag{B3}$$

By definition, stabilizer measurements amount to measuring the position and momentum operators $\hat{x}$ and $\hat{p}$ modulo $\sqrt{\pi}$. Clearly, all basis states and their superpositions have $\hat{x} = \hat{p} = 0$ modulo $\sqrt{\pi}$ and are stabilized by $\hat{S}_q$ and $\hat{S}_p$. As a result, any phase space shift error $\exp[i(\xi_p\hat{x} - \xi_x\hat{p})]$ acting on the GKP qubit can be detected and corrected as long as $|\xi_x|, |\xi_p| < \sqrt{\pi}/2$.

Pauli operators of the GKP qubit are given by the square root of the stabilizer generators as follows,

$$\hat{Z}_{\text{gkp}} = (\hat{S}_x)^{\frac{1}{2}} = \exp[i\sqrt{\pi}\hat{x}], \qquad \hat{X}_{\text{gkp}} = (\hat{S}_p)^{\frac{1}{2}} = \exp[-i\sqrt{\pi}\hat{p}], \tag{B4}$$

and it is easy to check that the action of the Pauli operators on the computational basis states is as expected,

$$\hat{Z}_{\text{gkp}} |0_{\text{gkp}}\rangle = |0_{\text{gkp}}\rangle, \qquad \hat{Z}_{\text{gkp}} |1_{\text{gkp}}\rangle = -|1_{\text{gkp}}\rangle,$$
$$\hat{X}_{\text{gkp}} |0_{\text{gkp}}\rangle = |1_{\text{gkp}}\rangle, \qquad \hat{X}_{\text{gkp}} |1_{\text{gkp}}\rangle = |0_{\text{gkp}}\rangle. \tag{B5}$$

An important property of GKP qubits is that the Clifford operations on the GKP qubits can be implemented using only Gaussian operations. More explicitly, generators of the Clifford group, $\hat{S}_{\text{gkp}}, \hat{H}_{\text{gkp}}$ and $\text{CNOT}_{\text{gkp}}^{j \to k}$ are given by

$$\hat{S}_{\text{gkp}} = \exp\left[i\frac{\hat{x}^2}{2}\right],$$
$$\hat{H}_{\text{gkp}} = \exp\left[i\frac{\pi}{2}\hat{a}^\dagger\hat{a}\right],$$
$$\text{CNOT}_{\text{gkp}}^{j \to k} = \exp[-i\hat{x}_j\hat{p}_k], \tag{B6}$$

Controlled-phase gate (which we need to prepare cluster states) can similarly be realized by Gaussian operations as

in

$$\text{CZ}_{\text{gkp}}^{j,k} = \exp[-i\hat{x}_j\hat{x}_k]. \tag{B7}$$

It is easy to see that

$$\begin{aligned}
\hat{S}_{\text{gkp}} |0_{\text{gkp}}\rangle &= |0_{\text{gkp}}\rangle, & \hat{S}_{\text{gkp}} |1_{\text{gkp}}\rangle &= i |1_{\text{gkp}}\rangle, \\
\hat{H}_{\text{gkp}} |0_{\text{gkp}}\rangle &= |+_{\text{gkp}}\rangle, & \hat{H}_{\text{gkp}} |1_{\text{gkp}}\rangle &= |-_{\text{gkp}}\rangle,
\end{aligned} \tag{B8}$$

and

$$\text{CNOT}_{\text{gkp}}^{j\to k} |\mu_{\text{gkp}}^{(j)}\rangle |\nu_{\text{gkp}}^{(k)}\rangle = |\mu_{\text{gkp}}^{(j)}\rangle |(\mu \oplus \nu)_{\text{gkp}}^{(k)}\rangle, \tag{B9}$$

$$\text{CZ}_{\text{gkp}}^{j,k} |\mu_{\text{gkp}}^{(j)}\rangle |\nu_{\text{gkp}}^{(k)}\rangle = e^{i\mu^{(j)}\nu^{(k)}} |\mu_{\text{gkp}}^{(j)}\rangle |\nu_{\text{gkp}}^{(k)}\rangle, \tag{B10}$$

where $|\mu_{\text{gkp}}^{(j)}\rangle := \sum_n |(2n+\mu)\sqrt{\pi}\rangle_{q_j}$ with $\mu,\nu = 0,1$ is the GKP state in the $j^{\text{th}}$ mode and $\oplus$ denotes the binary addition.

### 2. Error models for GKP qubits

As mentioned earlier, random shift errors to GKP qubits can be corrected. Such errors are usually modeled as a Gaussian random displacement error channel $\mathcal{G}[\sigma]$ given by

$$\mathcal{G}[\sigma](\hat{\rho}) := \int \frac{d^2\boldsymbol{\xi}}{2\pi\sigma^2} \exp\left[-\frac{|\boldsymbol{\xi}|^2}{2\sigma^2}\right] \hat{D}(\boldsymbol{\xi})\hat{\rho}\hat{D}^\dagger(\boldsymbol{\xi}), \tag{B11}$$

where $\hat{D}(\boldsymbol{\xi}) := \exp[i\xi_x\hat{p} - i\xi_p\hat{x}]$ is the displacement operator and $\boldsymbol{\xi} = (\xi_x, \xi_p) \in \mathbb{R}^2$ is the displacement vector. The error channel $\mathcal{G}[\sigma]$ randomly shifts the position and momentum quadratures by $\hat{x} \to \hat{x} + \xi_x$ and $\hat{p} \to \hat{p} + \xi_p$, where $\xi_x$ and $\xi_p$ follow a Gaussian random distribution with zero mean and standard deviation $\sigma$, i.e., $\xi_x, \xi_p \sim \mathcal{N}(0,\sigma)$. As we perform homodyne measurements, small random shifts can be detected while larger random shifts lead to Pauli errors. For instance, when measuring position (i.e., in $\hat{Z}_{\text{gkp}}$ basis) the small random position shifts $|\xi_x| < \sqrt{\pi}/2$ can be correctly identified, whereas the larger shifts $|\xi_x - \sqrt{\pi}| < \sqrt{\pi}/2$ are incorrectly identified as a smaller shift of $\xi_x - \sqrt{\pi}$. Such a misidentification implies a residual shift $\exp[-i\sqrt{\pi}\hat{p}] = \hat{X}_{\text{gkp}}$ and thus leads to a Pauli $\hat{X}_{\text{gkp}}$ error on the GKP qubit. Hence, assuming an ideal GKP qubit subject to the Gaussian noise model (B11) logical errors occur with probability

$$p_{\text{err}}(\sigma) = \frac{1}{\sqrt{2\pi\sigma^2}} \sum_{n\in\mathbb{Z}} \int_{(2n+1/2)\sqrt{\pi}}^{(2n+3/2)\sqrt{\pi}} e^{-z^2/2\sigma^2} dz. \tag{B12}$$

In the remainder of this appendix, we discuss the full error model of measurement-based computation based on GKP cluster states, including both preparation and execution errors. There are two sources of errors during the preparation: the finite squeezing of GKP qubits and the CZ gate error. There are also two sources of errors during the execution step, depending on the measurement basis. If the measurement is performed in the Pauli basis, we only incur an error associated with the finite efficiency of the homodyne measurement. However, measuring in a non-Pauli basis, e.g. $|0_{\text{gkp}}\rangle + e^{i\theta}|1_{\text{gkp}}\rangle$, involves applying a non-Clifford gate before the homodyne measurement. The non-Clifford gate usually requires extra ancilla GKP qubits and overwhelms the homodyne measurement error (as we discuss below). Hence, we consider only non-Clifford gate fidelity to evaluate non-Pauli basis measurement-flip error rates.

#### a. Approximate GKP states

Before we get into details of error models, let us recall that a realistic approximate GKP states can be modeled by applying a Gaussian envelope operator to an ideal GKP state $|\psi_{\text{gkp}}^\Delta\rangle \propto e^{-\Delta\hat{a}^\dagger\hat{a}}|\psi_{\text{gkp}}\rangle$. Expanding the envelope

operator in terms of displacement operators, we may write

$$|\psi_{\text{gkp}}^{\Delta}\rangle \propto \int \frac{d^2\boldsymbol{\xi}}{\pi} \text{Tr}\big[\exp[-\Delta \hat{n}]\hat{D}^{\dagger}(\boldsymbol{\xi})\big]\hat{D}(\boldsymbol{\xi})|\psi_{\text{gkp}}\rangle$$
$$\propto \int d^2\boldsymbol{\xi} \exp\Big[-\frac{|\boldsymbol{\xi}|^2}{4\sigma_{\text{gkp}}^2}\Big]\hat{D}(\boldsymbol{\xi})|\psi_{\text{gkp}}\rangle, \quad (B13)$$

where $\sigma_{\text{gkp}}^2 = (1-e^{-\Delta})/(1+e^{-\Delta}) \xrightarrow{\Delta \ll 1} \Delta/2$. In other words, an approximate GKP state can be understood as a coherent superposition of Gaussian peaks at ideal square lattice points (B1) along with a Gaussian envelope.

The effect of finite-squeezing in logical operations can be analyzed by Fock space simulations where the accuracy is controlled by the Fock space truncation number. As we approach the ideal limit $\Delta \ll 1$, more Fock space states need to be included in simulations, and thus calculations may quickly become unwieldy. For simplicity, we use an alternative approach to treat the finite-squeezing in terms of twirling approximation where a GKP state with finite squeezing is approximated by an incoherent mixture of random displacement errors as in

$$|0_{\text{gkp}}\rangle \to \mathcal{G}[\sigma_{\text{gkp}}](|0_{\text{gkp}}\rangle\langle 0_{\text{gkp}}|),$$
$$|+_{\text{gkp}}\rangle \to \mathcal{G}[\sigma_{\text{gkp}}](|+_{\text{gkp}}\rangle\langle +_{\text{gkp}}|) \quad (B14)$$

where we use the Gaussian random displacement error $\mathcal{G}[\sigma]$ defined in Eq. (B11). We should note that this method overestimates the error since we are dropping the coherent terms in Eq. (B13) [6]. Hence, the twirling approximation can be viewed as more pessimistic error model.

#### b. CZ-gate error

We model the CZ-gate error from first principles where the controlled-phase gate is implemented by evolving the system under the Hamiltonian $\hat{H} = g\hat{x}_1\hat{x}_2$ (where $g$ is the coupling strength) for $\Delta t = 1/g$ during which independent photon loss errors occur continuously in both input qubits. In other words, we replace the unitary time-evolution $\text{CZ}_{1,2} = \exp[i\hat{x}_1\hat{x}_2]$ by a completely positive and trace-preserving (CPTP) map $\exp[\mathcal{L}_{\text{cz}}\Delta t]$ where the Lindbladian generator $\mathcal{L}_{\text{cz}}$ is given by

$$\mathcal{L}_{\text{cz}}(\hat{\rho}) = -ig[\hat{x}_1\hat{x}_2, \hat{\rho}] + \kappa\big(\mathcal{D}[\hat{a}_1] + \mathcal{D}[\hat{a}_2]\big)\hat{\rho}. \quad (B15)$$

Here, $\mathcal{D}[\hat{A}](\hat{\rho}) := \hat{A}\hat{\rho}\hat{A}^{\dagger} - \frac{1}{2}\{\hat{A}^{\dagger}\hat{A}, \hat{\rho}\}$ and $\kappa$ is the photon loss rate.

Similar as before, to analytically derive the error model at the cost of overestimating it, we assume a noisier gate by adding heating errors $\kappa(\mathcal{D}[\hat{a}_1^{\dagger}] + \mathcal{D}[\hat{a}_2^{\dagger}])$ to the Lindbladian $\mathcal{L}_{\text{cz}}$, i.e.,

$$\mathcal{L}'_{\text{cz}} := \mathcal{L}_{\text{cz}} + \kappa\big(\mathcal{D}[\hat{a}_1^{\dagger}] + \mathcal{D}[\hat{a}_2^{\dagger}]\big), \quad (B16)$$

where the heating rate $\kappa$ is the same as the photon loss rate. This trick is used to convert the loss errors into random displacement errors (see Refs. [6, 7]). The resulting noisy CZ is equivalent to the ideal CZ followed by a correlated Gaussian random displacement error $\hat{x}_i \to \hat{x}_i + \xi_x^{(i)}$ and $\hat{p}_i \to \hat{p}_i + \xi_p^{(i)}$ for $i \in \{1, 2\}$, where the additive shift errors are drawn from bivariate Gaussian distributions $(\xi_x^{(1)}, \xi_p^{(2)}), (\xi_x^{(2)}, \xi_p^{(1)}) \sim \mathcal{N}(0, \Sigma_{\text{cz}})$ with the noise covariance matrix

$$\Sigma_{\text{cz}} = \sigma_{\text{cz}}^2 \begin{pmatrix} 1 & 1/2 \\ 1/2 & 4/3 \end{pmatrix}, \quad (B17)$$

and the variance $\sigma_{\text{cz}}^2$ is given by $\sigma_{\text{cz}}^2 = \kappa\Delta t = \kappa/g$.

In what follows, we derive the CZ gate error model in Eq. (B17) (using the method developed in Ref. [6]). The goal is to find the effective Gaussian channel associated with the time-evolution in Master equation with the Lindbladian $\mathcal{L}'_{\text{cz}} = \mathcal{V}_{\text{cz}} + \mathcal{L}_{\text{err}}$, where $\mathcal{V}_{\text{cz}}$ and $\mathcal{L}_{\text{err}}$ are defined as

$$\mathcal{V}_{\text{cz}}(\hat{\rho}) := ig[\hat{x}_1\hat{x}_2, \hat{\rho}],$$
$$\mathcal{L}_{\text{err}}(\hat{\rho}) := \kappa \sum_{k=1}^{2} \big(\mathcal{D}[\hat{a}_k] + \mathcal{D}[\hat{a}_k^{\dagger}]\big)\hat{\rho}. \quad (B18)$$

The time-evolution operator $\exp[\mathcal{L}'_{\text{cz}}\Delta t]$ with $\Delta t = 1/g$ can be evaluated using time-discretization and Trotter's formula,

$$\exp[\mathcal{L}'_{\text{cz}}\Delta t] = \lim_{N\to\infty}\left[\exp\left[\mathcal{V}_{\text{cz}}\frac{\Delta t}{N}\right]\exp\left[\mathcal{L}_{\text{err}}\frac{\Delta t}{N}\right]\right]^N. \tag{B19}$$

We note that both $\exp[\mathcal{V}_{\text{cz}}\Delta t/N]$ and $\exp[\mathcal{L}_{\text{err}}\Delta t/N]$ operations are Gaussian channels whose characteristic matrices are given by

$$\boldsymbol{T}_{\text{cz}} = \begin{pmatrix} 1 & 0 & 0 & 0 \\ 0 & 1 & 0 & 0 \\ 0 & 1/N & 1 & 0 \\ 1/N & 0 & 0 & 1 \end{pmatrix}, \quad \boldsymbol{N}_{\text{cz}} = 0,$$

$$\boldsymbol{T}_{\text{err}} = \begin{pmatrix} 1 & 0 & 0 & 0 \\ 0 & 1 & 0 & 0 \\ 0 & 0 & 1 & 0 \\ 0 & 0 & 0 & 1 \end{pmatrix}, \quad \boldsymbol{N}_{\text{err}} = \frac{\kappa\Delta t}{N}\begin{pmatrix} 1 & 0 & 0 & 0 \\ 0 & 1 & 0 & 0 \\ 0 & 0 & 1 & 0 \\ 0 & 0 & 0 & 1 \end{pmatrix}, \tag{B20}$$

respectively. We recall that a Gaussian channel characterized by two matrices $\boldsymbol{T}$ (generator of the symplectic transformation) and $\boldsymbol{N}$ (noise matrix) acting on a Gaussian state with covariance matrix $\boldsymbol{\Sigma}$ implements the following transformation [8],

$$\boldsymbol{\Sigma} \to \boldsymbol{T}\boldsymbol{\Sigma}\boldsymbol{T}^T + \boldsymbol{N}. \tag{B21}$$

In our case, the quadrature operator $\hat{\boldsymbol{r}} = (\hat{x}_1, \hat{x}_2, \hat{p}_1, \hat{p}_2)^T$ is transformed according to the following symplectic transformation

$$\hat{\boldsymbol{r}} \to (\boldsymbol{T}_{\text{cz}})^N \hat{\boldsymbol{r}} = \begin{pmatrix} 1 & 0 & 0 & 0 \\ 0 & 1 & 0 & 0 \\ 0 & 1 & 1 & 0 \\ 1 & 0 & 0 & 1 \end{pmatrix}\hat{\boldsymbol{r}} = \begin{pmatrix} \hat{x}_1 \\ \hat{x}_2 \\ \hat{p}_1 + \hat{x}_2 \\ \hat{p}_2 + \hat{x}_1 \end{pmatrix}, \tag{B22}$$

as expected for a CZ gate. Furthermore, the covariance matrix $\boldsymbol{\Sigma}$ is modified accordingly

$$\boldsymbol{\Sigma} \to (\boldsymbol{T}_{\text{cz}})^N \boldsymbol{\Sigma}((\boldsymbol{T}_{\text{cz}})^N)^T + \sum_{k=1}^{N}(\boldsymbol{T}_{\text{cz}})^k \boldsymbol{N}_{\text{err}}(\boldsymbol{T}_{\text{cz}}^T)^k$$

$$= (\boldsymbol{T}_{\text{cz}})^N \boldsymbol{\Sigma}((\boldsymbol{T}_{\text{cz}})^N)^T + \sum_{k=1}^{N}\frac{\kappa\Delta t}{N}\begin{pmatrix} 1 & 0 & 0 & \frac{k}{N} \\ 0 & 1 & \frac{k}{N} & 0 \\ 0 & \frac{k}{N} & 1+(\frac{k}{N})^2 & 0 \\ \frac{k}{N} & 0 & 0 & 1+(\frac{k}{N})^2 \end{pmatrix}$$

$$= (\boldsymbol{T}_{\text{cz}})^N \boldsymbol{\Sigma}((\boldsymbol{T}_{\text{cz}})^N)^T + \kappa\Delta t\begin{pmatrix} 1 & 0 & 0 & 1/2 \\ 0 & 1 & 1/2 & 0 \\ 0 & 1/2 & 4/3 & 0 \\ 1/2 & 0 & 0 & 4/3 \end{pmatrix}, \tag{B23}$$

where in the last line the limit $N \to \infty$ is taken. Therefore, the noisy controlled-phase gate can be understood as the ideal gate followed by a correlated Gaussian random displacement error with the covariance matrix $\Sigma_{\text{cz}}$ given in Eq. (B17).

### c. Arbitrary single-qubit rotation gate error

Clifford gates on GKP qubits can be realized using Gaussian operations, while the implementation of non-Clifford gates requires non-Gaussian operations. As mentioned in the main text, two types of approaches have been devised to implement the T-gate, a non-Clifford operation sufficient for a universal GKP gate set. One approach is to use the cubic phase gate combined with Gaussian operations. The other approach is based on a magic-state injection method and a quantum nondemolition (QND) gate. In this approach, a non-Gaussian ancillary state is used together with Clifford operations and measurements. The former approach typically has a low gate fidelity due to inefficacy of

cubic phase gate which only involves the lowest order non-Gaussian operation. Also, the direct implementation of the nondemolition gate in the latter approach suffers from photon loss. To address both issues, a measurement-induced T-gate based on the magic-state injection method has been developed and demonstrated in experiments [9–11], which was further improved recently [12].

The non-Pauli measurement basis required for MBQC is an arbitrary vector in $xy$-plane characterized by an angle $0 \leq \theta \leq \pi$. This measurement basis can be realized by an arbitrary rotation around $z$-axis

$$\hat{U}_{Z_{\text{gkp}}}(\theta) = |0_{\text{gkp}}\rangle \langle 0_{\text{gkp}}| + e^{i\theta} |1_{\text{gkp}}\rangle \langle 1_{\text{gkp}}|, \tag{B24}$$

combined with a Hadamard gate. To this end, we imagine using either a cubic phase gate or the measurement-induced magic-state injection, where the offline magic state $|0_{\text{gkp}}\rangle + e^{i\pi/4}|1_{\text{gkp}}\rangle$ is replaced by $|0_{\text{gkp}}\rangle + e^{i\theta}|1_{\text{gkp}}\rangle$.

The infidelity of non-Clifford gates can be modeled by an ideal gate accompanied by a depolarizing channel

$$\varepsilon_a(\hat{\rho}) = (1 - p_t(\sigma_{\text{in}}))\hat{\rho} + \frac{p_t(\sigma_{\text{in}})}{3} \sum_{P_a \in \mathcal{P}_{1,\text{gkp}}} \hat{P}_{a,\text{gkp}} \hat{\rho} \hat{P}_{a,\text{gkp}} \tag{B25}$$

where $a$ labels the noisy GKP state with $\sigma_{\text{in}}$ (i.e., an ideal GKP state with a random displacement error of variance $\sigma_{\text{in}}^2$), the Pauli operators are defined in Eq. (B5) and error probability $p_t(\sigma_{\text{in}})$ is related to average gate fidelity via $p_t(\sigma_{\text{in}}) = \frac{3}{2}[1 - \overline{\mathcal{F}}(\sigma_{\text{in}})]$ which depends on the input squeezing level $\sigma_{\text{in}}$. Assuming another Gaussian error channel at the output (e.g., due to measurement efficiency, etc.) which is modeled by a random displacement with variance $\sigma_{\text{out}}^2 = \sigma_{\text{gkp}}^2 + \sigma_m^2$, the combined error model $\varepsilon_a' = \mathcal{G}_a[\sigma_{\text{out}}] \circ \varepsilon_a$ is a Pauli error channel defined as

$$\varepsilon_a'(\hat{\rho}) = (1 - p_t(\sigma_{\text{in}}))\mathcal{G}_a[\sigma_{\text{out}}](\hat{\rho}) + \frac{p_t(\sigma_{\text{in}})}{3} \sum_{P_a \in \{X,Y,Z\}} \mathcal{G}_a[\sigma_{\text{out}}]\left(\hat{P}_{a,\text{gkp}} \hat{\rho} \hat{P}_{a,\text{gkp}}\right). \tag{B26}$$

Hence, the effective measurement-flip error rate is given by

$$\begin{aligned} e_{nc} &= (1 - p_t(\sigma_{\text{in}})) p_{\text{err}}(\sigma_{\text{out}}) + \frac{2 p_t(\sigma_{\text{in}})}{3} (1 - p_{\text{err}}(\sigma_{\text{out}})) \\ &\approx \frac{2 p_t(\sigma_{\text{in}})}{3} + p_{\text{err}}(\sigma_{\text{out}}). \end{aligned} \tag{B27}$$

($nc$: non-Clifford) where the Gaussian error rate is defined in Eq. (B12). The above approximation is justified since $p_{\text{err}} \ll p_t$ for experimentally relevant systems.

### 3. Error model for MBQC on GKP cluster states

To model the cluster state initialization error, we start with $N$ approximate GKP qubits (B14), which can be written in a more compact form as

$$\begin{aligned} \hat{\rho}_0 &= \mathcal{G}_{\boldsymbol{\Sigma}_0}(|+_{\text{gkp}}\rangle \langle +_{\text{gkp}}|^{\otimes N}) \\ &\propto \int d^{2N}\boldsymbol{\xi} \exp\left[-\frac{1}{2}\boldsymbol{\xi}^T \boldsymbol{\Sigma}_0^{-1} \boldsymbol{\xi}\right] \hat{D}(\boldsymbol{\xi})(|+\rangle \langle +|_{\text{gkp}}^{\otimes N})\hat{D}^\dagger(\boldsymbol{\xi}), \end{aligned} \tag{B28}$$

where $\boldsymbol{\xi} = (\boldsymbol{\xi_x}, \boldsymbol{\xi_p})^T = (\xi_{x_1}, \cdots, \xi_{x_N}, \xi_{p_1}, \cdots, \xi_{p_N})^T$, and $\hat{\boldsymbol{r}} = (\hat{x}_1, ..., \hat{x}_N, \hat{p}_1, ..., \hat{p}_N)^T$ associated with the $N$ modes. Here, $\boldsymbol{\Sigma}_0 \geq 0$ is the Gaussian covariance matrix where we have $\boldsymbol{\Sigma}_0 = \sigma_{\text{gkp}}^2 \mathbb{1}$ for the initial state. Furthermore, the multi-mode displacement operator is defined as $\hat{D}(\boldsymbol{\xi}) = \exp[i\boldsymbol{\xi}^T \boldsymbol{\Omega} \hat{\boldsymbol{r}}]$, and $\boldsymbol{\Omega}$ is the anti-symmetric symplectic metric.

Next, we apply CZ gates to the neighboring qubits connected by an edge on the state graph. This is implemented by conjugating with the corresponding unitary operator $\hat{U}_C$,

$$\begin{aligned} \hat{\rho}_{\text{cs}} &= \hat{U}_{\text{cs}} \hat{\rho}_0 \hat{U}_{\text{cs}}^\dagger \\ &= \mathcal{G}_{\boldsymbol{\Sigma}}\left[\left(\hat{U}_{\text{cs}} |+_{\text{gkp}}\rangle \langle +_{\text{gkp}}|^{\otimes N} \hat{U}_{\text{cs}}\right)\right]. \end{aligned} \tag{B29}$$

where exchanging the unitary evolution and the noise channel amounts to a symplectic transformation of the noise

covariance matrix as

$$\mathbf{\Sigma}_0 \to \mathbf{\Sigma} = \mathbf{S}_C \mathbf{\Sigma}_0 \mathbf{S}_C^T + \widetilde{\mathbf{\Sigma}}_{\text{cz}}, \tag{B30}$$

and

$$\mathbf{S}_C = \begin{pmatrix} \mathbb{1} & \mathbf{0} \\ \mathbf{A} & \mathbb{1} \end{pmatrix}. \tag{B31}$$

Here, $\mathbf{A}$ is a $N \times N$ adjacency matrix of the cluster state, i.e., $(i,j)$ entry is 1 if two nodes are entangled with a CZ gate and 0 otherwise. This is because applying a CZ gate is a Gaussian operation, and the resulting state remains Gaussian where the new covariance matrix is related to the initial one via the corresponding symplectic transformation. We note that noisy CZ gates contribute an additional Gaussian noise with the covariance matrix $\Sigma_{\text{cz}}$ of Eq. (B17). The CZ gates are applied sequentially and this error is propagated to the next qubit through the next CZ gate (we use $\widetilde{\mathbf{\Sigma}}_{\text{cz}}$ to differentiate the effective covariance matrix with Eq. (B17)). As a result the diagonal component of the covariance matrix for the momentum quadrature is modified to $\frac{11}{3}\sigma_c^2$ as opposed to $2 \times \frac{4}{3}\sigma_c^2$.

Lastly, the homodyne measurement efficiency adds to the covariance matrix as in

$$\mathbf{\Sigma} \to \mathbf{\Sigma} + \sigma_m^2 \mathbb{1}, \tag{B32}$$

where we use the twirling approximation for the variance of inefficient measurement. Next, we explain how to calculate the effective error channel of quantum gates implemented on cluster states.

To numerically compute the measurement error rates of noisy GKP states in a cluster state, we use Monte Carlo simulations where Gaussian random displacement vectors $\boldsymbol{\xi}$ are generated according to the final covariance matrix (B32) and depending on the measurement basis of each qubit we determine an error event by checking whether the corresponding entry in the displacement vector is greater than $\sqrt{\pi}/2$ or not.

Assuming that correlated errors due to the non-diagonal elements in the covariance matrix are not large compared to variances of individual modes, we can evaluate the error probability of homodyne measurement outcomes by the marginal probability of each mode separately. The reduced covariance matrix of each mode is given by

$$\mathbf{\Sigma}_a = \begin{pmatrix} \sigma_q^2 & 0 \\ 0 & \sigma_p^2 \end{pmatrix}, \tag{B33}$$

where

$$\sigma_q^2 = \sigma_{\text{gkp}}^2 + 2\sigma_c^2 + \sigma_m^2, \tag{B34}$$

$$\sigma_p^2 = 3\sigma_{\text{gkp}}^2 + \frac{11}{3}\sigma_c^2 + \sigma_m^2. \tag{B35}$$

This leads to the measurement error probability calculated by Eq. (B12) where we plug in the corresponding diagonal entries of the above covariance matrix. More explicitly, the measurement-flip error probability associated with measuring qubit $a$ is given by

$$\begin{aligned} e_{X_{\text{gkp}}} &= p_{\text{err}}(\sigma_p), \\ e_{Y_{\text{gkp}}} &= p_{\text{err}}([(\sigma_q^2 + \sigma_p^2)/2]^{1/2}), \\ e_{Z_{\text{gkp}}} &= p_{\text{err}}(\sigma_q), \end{aligned} \tag{B36}$$

where the subscript indicates the Pauli measurement basis and the error function $p_{\text{err}}(\cdot)$ is defined in Eq. (B12).

For non-Pauli basis measurements, we use an estimated error rate [12] as a function of the input variance defined by

$$p_t(\sigma) = p_{\text{err}}(\sqrt{k}\sigma), \tag{B37}$$

where $k > 1$ is a free parameter which we choose such that the infidelity of non-Clifford measurements would be an order of magnitude worse than that of Pauli measurements. As the input variance $\sigma_{\text{in}}$, we use the mean value of variances in Eq. (B34) with $\sigma_m = 0$.

#### 4. Effective gate fidelity

In this part, we use the measurement-flip error probabilities calculated in the previous part to evaluate the effective error rates of the measurement-based gates implemented on noisy cluster states.

*a. Identity gate*

The identity gate is realized on a linear cluster of three qubits by the measurement pattern $\mathcal{M}_{X_2}\mathcal{M}_{X_1}$. The overall error is a Pauli error channel with

$$\begin{aligned} q_x &= e_2(1-e_1), \\ q_y &= e_1 e_2, \\ q_z &= e_1(1-e_2), \end{aligned} \quad (B38)$$

where $e_1 = e_2 = e_{X_{\text{gkp}}}$. The averaged gate fidelity is found to be

$$\begin{aligned} \overline{F}(\mathcal{M}_e, I) &= 1 - \frac{2}{3}(e_1 + e_2 - e_1 e_2) \\ &= 1 - \frac{2}{3}(2e_{X_{\text{gkp}}} - e_{X_{\text{gkp}}}^2). \end{aligned} \quad (B39)$$

*b. Hadamard*

The hadamard gate is implemented on a linear cluster of five qubits by the measurement pattern $\mathcal{M}_{Y_4}\mathcal{M}_{Y_3}\mathcal{M}_{Y_2}\mathcal{M}_{X_1}$. Because of the choice of measurement basis we have $e_1 = e_{X_{\text{gkp}}}$ and $e_2 = e_3 = e_4 = e_{Y_{\text{gkp}}}$. The overall error is again a Pauli error channel where the probability coefficients are given by

$$\begin{aligned} q_x &= e_{X_{\text{gkp}}}(1-e_{Y_{\text{gkp}}})^3 + e_{Y_{\text{gkp}}}(1-e_{X_{\text{gkp}}})(1-e_{Y_{\text{gkp}}})^2 + e_{Y_{\text{gkp}}}^2(1-e_{X_{\text{gkp}}})(1-e_{Y_{\text{gkp}}}) + e_{X_{\text{gkp}}}^3, \\ q_y &= q_z = e_{Y_{\text{gkp}}}(1-e_{Y_{\text{gkp}}})^2 + e_{Y_{\text{gkp}}}^2(1-e_{Y_{\text{gkp}}}). \end{aligned} \quad (B40)$$

The gate fidelity to linear order in $e_i$ is found to be

$$\overline{F}(\mathcal{M}_e, H) = 1 - \frac{2}{3}(e_{X_{\text{gkp}}}^2 + 3e_{Y_{\text{gkp}}}^2). \quad (B41)$$

*c. Phase gate*

The phase gate is implemented on a linear cluster of five qubits by the measurement pattern $\mathcal{M}_{X_4}\mathcal{M}_{Y_3}\mathcal{M}_{X_2}\mathcal{M}_{X_1}$. Because of the choice of measurement basis this time we have $e_1 = e_2 = e_4 = e_{X_{\text{gkp}}}$ and $e_3 = e_{Y_{\text{gkp}}}$. Similarly, the overall error is a Pauli error channel where the probability coefficients are given by

$$\begin{aligned} q_x = q_y &= e_{X_{\text{gkp}}}(1-e_{X_{\text{gkp}}}), \\ q_z &= e_{X_{\text{gkp}}} + e_{Y_{\text{gkp}}} - 4e_{X_{\text{gkp}}}e_{Y_{\text{gkp}}} - e_{X_{\text{gkp}}}^2(1-4e_{Y_{\text{gkp}}}). \end{aligned} \quad (B42)$$

The gate fidelity to linear order in $e_i$ is found to be

$$\overline{F}(\mathcal{M}_e, S) = 1 - \frac{2}{3}(e_{Y_{\text{gkp}}}^2 + 3e_{X_{\text{gkp}}}^2). \quad (B43)$$

*d. U3 gate*

Non-Clifford measurements are essential to arbitrary single qubit rotations. Here, as we discussed for the depolarizing channel in Appendix A 1 b, we are ignoring angle-dependent terms due to adaptive measurement basis since

they are of higher orders.

$$q_x = 2e_{nc}, \qquad q_y = O(e^2), \qquad q_z = e_{X_{\mathrm{gkp}}} + e_{nc}. \tag{B44}$$

In our simulations we set $k = 5$ in Eq. (B37) to evaluate $e_{nc}$ (B27).

---



\end{document}